# Anti-aliased metasurfaces beyond the Nyquist limit

Seokwoo Kim[1]†, Joohoon Kim[1]†, Kyungtae Kim[1]†, Minsu Jeong[1], and Junsuk Rho[1,2,3,4,5]*

[1]Department of Mechanical Engineering, Pohang University of Science and Technology (POSTECH), Pohang 37673, Republic of Korea

[2]Department of Chemical Engineering, Pohang University of Science and Technology (POSTECH), Pohang, 37673, Republic of Korea

[3]Department of Electrical Engineering, Pohang University of Science and Technology (POSTECH), Pohang 37673, Republic of Korea

[4]POSCO-POSTECH-RIST Convergence Research Center for Flat Optics and Metaphotonics, Pohang 37673, Republic of Korea

[5]National Institute of Nanomaterials Technology (NINT), Pohang 37673, Republic of Korea

†These authors contributed equally to this work.

*Corresponding author. Email: jsrho@postech.ac.kr

## Abstract

Sampling is a pivotal element in the design of metasurfaces, enabling a broad spectrum of applications. Despite its flexibility, sampling can result in reduced efficiency and unintended diffractions, which are more pronounced at high numerical aperture or shorter wavelengths, *e.g.* ultraviolet spectrum. Prevailing metasurface research has often relied on the conventional Nyquist sampling theorem to assess sampling appropriateness, however, our findings reveal that the Nyquist criterion is insufficient for preventing the diffractive distortion. Specifically, we find that the performance of a metasurface is significantly correlated to the geometric relationship between the spectrum morphology and sampling lattice. Based on lattice-based diffraction analysis, we demonstrate several anti-aliasing strategies from visible to ultraviolet regimes. These approaches significantly reduce aliasing phenomena occurring in high numerical aperture metasurfaces.

Metasurfaces are flat optical components that have drawn substantial attention for their various abilities to control electromagnetic waves. Among these, phase gradient metasurfaces, which manipulate the wavefront of light at a subwavelength scale, stand out due to their versatility (*1-3*). Wavefront transformation is achieved by periodically arranged subwavelength scatterers, so-called meta-atoms. They can encode the local phase within each pixel according to the desired phase profile. When incident waves pass through the metasurface, the desired wavefront becomes reconstructed. This sampling-based design principle can be applied to various applications, *e.g.* lenses (*4-6*), holograms (*7-12*), biosensors (*13-15*), vortex beams (*16,17*), and structured light (*18,19*).

However, sampling is a double-edged sword. Despite the flexibility, metasurfaces encoded by high numerical aperture (NA) phase function suffer from fundamental efficiency limits (*20,21*). Furthermore, coarse sampling leads to unintended diffractions having distorted momenta (*22,23*), which can be understood as a wagon-wheel effect in the metasurface (Fig. 1). These issues hinder the metasurface from high NA operation with high signal-to-noise ratio (SNR), which is crucial in extending detection ranges in LiDAR systems (*18,19,24*), broadening the field of view in augmented reality/virtual reality (AR/VR) devices (*25,26*), and enhancing resolution in imaging and fabrication systems (*27,28*). Fine sampling through smaller pixel size can alleviate those challenges, however, it leads to decreased effective index or reduced anisotropy of meta-atoms, which demands a high aspect ratio for full phase modulation, making fabrication more challenging (*29,30*). Furthermore, a small area of unit cells limits the degree of freedom in designing complex meta-atoms (*31,32*). As the wavelength shortens, especially in the ultraviolet (UV) regime, sampling problems become severe due to the more stringent sampling condition and relatively low refractive index of a lossless dielectric in the UV spectrum (*33*). Nevertheless, not only an analysis of distorted diffractions but also the anti-aliasing strategies have yet to be fully studied.

Here, we investigate the aliasing phenomena occurring in metasurfaces. We find that the higher-order spectra encoded in metasurfaces can be outcoupled into free space, manifesting as the wagon-wheel effect in diffraction. Moreover, the Nyquist sampling criterion, which is commonly employed in assessing sampling appropriateness, cannot fully determine the exact condition for avoiding distorted diffraction. Through expanding the diffraction analysis into a two-dimensional spectral space, we find that the geometric relationship between the phase spectrum and lattice structure plays a crucial role in determining the aliasing of the metasurface. Based on insights obtained from aliasing analysis, we propose anti-aliasing strategies to eliminate distorted diffraction. These approaches markedly eliminate aliasing phenomena occurring in high NA metasurfaces for visible to near-UV metasurfaces.

**Diffraction regimes of metasurfaces**

First, we start with the simplest case, a one-dimensional (1D) phase gradient metasurface for monochromatic light ($\lambda = \lambda_0$). The meta-atoms are arranged along the $x$-axis with a subwavelength interval ($\Lambda$) to sample the desired phase profile (Fig. 2A). Discrete sampling of the phase profile reproduces the periodic array of the spatial spectrum with interval $G = 2\pi/\Lambda$ in the momentum space (*35*). Among them, the higher-order spectra have shifted momenta by $nG$ for an integer $n$. The momenta $k$ from higher-order spectra located inside the light cone (*i.e.* $|k| < k_0$ where $k_0 = 2\pi/\lambda_0$) can propagate into the free space, manifesting as an aliasing of metasurface (Supplementary text section "Aliasing in 1D phase gradient metasurface"). To identify the spectral range where aliasing occurs, we construct the 'diffraction diagram' from repetitive light cones in the reciprocal space (Supplementary text section "Constructing a diffraction diagram by repetitive light cones"). Here, the light-cone array is arranged with interval $G$ in the momentum space (Fig. 2A, B). One can imagine that the light cones become closer together as the shorter wavelength ($\lambda_0$) and the wider sampling interval ($\Lambda$). Especially, when the parameter $\lambda_0/\Lambda$ is under the value of 2, light cones start to overlap each other. For any momentum $k_{meta}$ in the desired spectrum, it can be coupled to a higher-order light cone when it belongs to an overlap spectral region (Fig. 2B). It results in the aliasing of the metasurface. The parameter $\lambda_0/\Lambda$ thus can be regarded as a sampling level of the metasurface since it determines the spectral range susceptible to distortion.

Here, we consider the metasurface where light cones overlap each other, *i.e.* $\lambda_0/\Lambda < 2$. Assumes an arbitrary momentum $k_{meta} > 0$ within the desired spectrum (Fig. 2B). Depending on the magnitude of $k_{meta}$, there are three distinct diffraction scenarios for 1D metasurface:

**Aliasing-free (AF) regime** - When $k_{meta}$ is in a non-overlap spectral region, *i.e.* $|k_{meta}| < G - k_0$, only the $0^{th}$-order light cone is coupled. Therefore, the ideal order $T_{Ideal}$ radiates to the free space whereas the aliasing order $T_{Aliasing}$ doesn't appear. This aliasing-free spectral region is colored yellow in Fig. 2B.

**Weak wagon-wheel effect (WW) regime** - If $k_{meta}$ is within the range where light cones overlap but still inside the Brillouin zone (BZ), *i.e.* $G - k_0 < |k_{meta}| < G/2$, the incident light is coupled to both $0^{th}$-order and adjacent higher-order light cones. Different from the AF regime, the coupling of higher-order light cones leads to the coupling of aliasing order $k_{aliasing} = k_{meta} - G$ whose size of imparted transverse momentum is larger than those of ideal order. Note that the spectrum belonging to the WW regime satisfies the Nyquist criterion since the BZ boundary corresponds to the Nyquist limit ($|k_{meta}| = G/2$). This implies that, counterintuitively, the Nyquist criterion cannot ensure unintended diffraction in the free space. This spectral region is colored red in Fig. 2B.

**Strong wagon-wheel effect (SW) regime** – When $k_{meta}$ lies outside the BZ, *i.e.* $G/2 < |k_{meta}|$, incident light is coupled to both $0^{th}$-order and adjacent higher-order light cones where the transverse momentum of aliasing order is smaller than those of ideal order, *i.e.* $|k_{meta} - G| < |k_{meta}|$. The spectrum belonging to this regime does not satisfy the Nyquist sampling criterion since it is outside of BZ. This spectral region is colored blue in Fig. 2B.

As a proof of concept, the diffraction efficiencies of the beam steering metasurface are calculated for various sampling levels (Fig. 2C-E). The metasurface is targeted for normally incident light with a wavelength of 325 nm. The rigorous coupled-wave analysis (RCWA) is employed to determine diffraction efficiencies of various steering angles (see Materials and Methods). Different sets of meta-atoms are used for each sampling level whose refractive index is 2, and their height is determined to ensure full phase modulation from 0 to $2\pi$ (fig. S3). The red and blue curves in Fig. 2C-E represent the diffraction efficiency of the ideal order ($T_{Ideal}$) and aliasing ($T_{Aliasing}$), respectively. Here, we define diffraction efficiency as the ratio of power coupled into each diffraction order to the total transmitted light.

In Fig. 2C-E, $T_{Ideal}$ decreases as the steering angle increases, due to the reduced number of sampled phases since the phase gradient becomes steeper. When the light cones overlap, *i.e.* the sampling level is under 2, the WW and SW regimes emerge (Fig. 2D-E). At the WW regime, aliasing starts to emerge, and its efficiency increases as the steering angle increases. This is because the magnitude of imparted momentum of aliasing order $|k_{meta} - G|$ decreases as the desired momentum $k_{meta}$ increases. In the SW regime, the efficiency of the aliasing order eventually surpasses the ideal order. Note that the efficiency of the ideal (aliasing) order for $k_{meta} = G/2-\Delta k$ is equal to the efficiency of aliasing (ideal) for $k_{meta} = G/2+\Delta k$ by inversion symmetry of the phase profile (fig. S4). This implies that for metasurfaces where aliasing exists, attempts to increase the diffraction efficiency can concurrently amplify the aliasing effect.

**Anisotropic performance of sampling lattice**

Now, the aliasing analysis is further extended to a metasurface composed of a two-dimensional (2D) lattice, which is a more general case (fig. S5). While square lattice and hexagonal lattice are commonly employed in metasurfaces, it is possible to generate an arbitrary sampling lattice using two nonparallel vectors (*35*). Akin to a previous 1D case, we can construct a diffraction diagram for identifying diffraction regimes from the array of light cones positioned at each point generated by the linear combination of reciprocal basis vectors (Fig. 3A-B). For instance, figure 3C presents the diffraction diagram for a square lattice at a sampling level of 1.58 where yellow, red, and blue colored spectral regions in the diagram represent AF, WW, and SW diffraction regimes, respectively.

Consider the scenario where the arbitrary momentum $\vec{k}_{meta}$ is imparted to incident light through this sampling lattice. Here, NA for $\vec{k}_{meta}$ is assumed to be 0.82, which is represented by the red dashed circle (Fig. 3C). The sampling lattice doesn't satisfy the Nyquist criterion (NA < $\lambda_0/2\Lambda$ = 0.79). Interestingly, one can observe that the diffraction regimes vary depending on the azimuthal direction of $\vec{k}_{meta}$. Specifically, if the direction of the momentum is parallel to the lattice vectors, along the $x$ or $y$-axis, then the momentum falls into the SW region, however, if the momentum's direction is diagonal, the spectrum then belongs to the AF regime (Fig. 3C). This observation implies that each sampling lattice has specific directions that are particularly advantages for large-angle diffraction, which is not observed in the previous 1D case. Moreover, simply aligning the superior direction of the sampling lattice with that of the directional spectrum can serve as an effective anti-aliasing strategy.

As a proof of concept, we experimentally demonstrate the Pancharatnam-Berry metasurface (*1,36*) designed for steering light at an angle of 55°. The metasurface is designed based on a square lattice with a periodicity of 400 nm, targeting a wavelength of 632 nm, which corresponds to a sampling level of 1.58. For the meta-atoms, hydrogenated amorphous silicon (a-Si:H) is employed, with a refractive index of 2.42 at the target wavelength (see Materials and Methods and fig. S6). To validate the anisotropic performance of the metasurface, we fabricate ten different metasurfaces with identical phase profiles, but different sampling grids rotated from 0° to 45° with 5° intervals (Fig. 3D and scanning electron microscope (SEM) images are in fig. S7). Figure 3E shows the efficiency ratio between ideal and aliasing diffraction order measured by a power meter. Blue, red, and yellow colors denote the diffraction regimes obtained from the diffraction diagram represented in Fig. 3C. When the rotation angle is below about 15°, the diffraction belongs to the SW regime where the efficiency of aliasing slightly surpasses those of ideal order. When the rotation angle exceeds about 33°, aliasing nearly disappears even though the periodicity of the lattice doesn't meet the Nyquist criterion. Figure 3F shows the momentum space images compared with simulated results (Optical setup: fig. S8). One can find that while the direction of the ideal order is invariant due to the identical phase profile for metasurfaces, the aliasing order moves toward the outer edge of the light cone as the rotation angle increases. It is because the transverse momentum of the aliasing order ($\vec{k}_{\text{meta}}-\vec{v}_2$) varies according to the grid rotation angle, which affects the direction of $\vec{v}_2$. At the rotation angle above 30°, the aliasing order moves beyond the NA of the objective lens (0.9) and is no longer observable (fig. S9).

**Lattice-dependent characteristics in metalens**

Now we investigate the effect of sampling lattice in a metalens, which is one of the prominent applications for phase-gradient metasurfaces. Unlike the directional momentum encoded for beam steering, the spectrum for beam focusing is isotropic for all azimuthal directions. In addition, for a high NA metalens, the spectrum nearly spans the entire cross-section of a light cone. Consequently, high NA metalenses can exhibit varying diffraction regimes depending on the radial distance from the center. For example, for the high NA metalens having a low sampling level, aliasing can prevail at the edge of the metalens. Thus, the light transmitted through the high NA metalens mostly coupled to the focusing ideal orders, diverging aliasing orders, and non-diffracted zeroth orders (Fig. 4A).

Figure 4B presents a diffraction diagram for a hexagonal lattice and a square lattice. Although two lattices have an identical sampling level (1.26), they exhibit distinct shapes and areas for diffraction regimes (fig. S10). The range of NA of metalens for avoiding the SW regime, *i.e.* the Nyquist criterion, becomes NA $< \frac{\lambda}{\sqrt{3}\Lambda} \approx 0.58\frac{\lambda}{\Lambda}$ for a hexagonal lattice. In contrast, the coefficient on the right-hand side of the equality becomes 0.5 for the square lattice case, indicating a more stringent condition compared to a hexagonal lattice (fig. S11).

As a proof of concept, we fabricate the hexagonal lattice and square lattice metalenses for NA of 0.99 and 0.65, respectively (see Materials and Methods and fig. S12). All metalens are based on a hyperbolic phase profile and have the same periodicity (500 nm), target wavelength (632 nm), and sampling level of 1.26. The meta-atom table for square lattice and hexagonal lattice is in fig. S13. Figure 4C-F shows the light intensity distribution after passing through metalenses for NA of 0.99 and 0.65, respectively (Optical setup: fig. S14). Metalenses sampled on different lattices show distinct diffraction patterns, particularly associated with aliasing radiating outward to the optical axis. When NA approaches unity, the spectrum spans the almost entire light cone so that clear aliasing patterns are observed (see Fig 4C-D and aliasing patterns for various NA metalens are in Movie 1). When NA is 0.65, almost no aliasing is observed in the hexagonal lattice metalens, whereas the square lattice metalens exhibit aliasing channels radiating to four orthogonal directions due to coupling with adjacent light cones. The calculated focusing efficiencies for the hexagonal and square lattice metalenses (NA = 0.65) are 49.3% and 40.2%, respectively where the focusing efficiency is defined as the ratio of incoming power in the circle encircling with a radius of 3 times full-width half maximum of PSF to the input power. The quantity of unwanted diffraction for two different lattice metalenses, which can be obtained as the difference between the transmittance and the focusing efficiency, is 18.4% and 26.3%, respectively. The diffraction diagram in Fig. 4B reveals the existence of partial spectra coupling with the adjacent light cones even in hexagonal lattice metalenses, but these are nearly unobserved since the NA of the objective lens used in measurements limits the observable spectrum in the momentum space (fig. S15).

**Anti-aliasing by integrated lattice**

The resolution of standard lithography techniques constrains the feasible dimensions for unit cells of metasurface. Consequently, the sampling level of a metasurface inevitably decreases with a shorter wavelength, leading to more pronounced sampling-induced issues in the UV spectrum application. To mitigate this, we counteract aliasing phenomena in the high NA (0.9) metalens targeted for near-UV light (405 nm), by integrating an array of meta-dimers on the periphery of the metalens (Fig. 5D). The meta-atom table for UV metalens is in fig. S16. Each meta-dimer consists of two closely located nanostructures (Fig. 5A). The small intervening gap facilitates non-local coupling between two vertical waveguides. This coupling results in directional energy flows (*37*) (its time-averaged Poynting vector representation is depicted in Fig. 5B) within the meta-dimer. This resulted from the 'beating effect' that is interference between several coupled guided modes, rendering it more effective for large-angle scattering compared to conventional sampling-based metasurfaces (*27,37-39*).

The integration of meta-dimers has been optimized for various steering angles using RCWA (table S1). Notably, the diffraction efficiency of the meta-dimer array exceeds that of conventional metasurfaces at an angle of about 24° (fig. S17). Beyond this transition angle, we combine the meta-dimer array with the hexagonal lattice metalens, taking into account the phase of light diffracted after transmission through the meta-dimers, as shown in Fig. 5D. The meta-dimer array is strategically positioned along the peripheries of concentric circles, with each circle's radius tailored to align the meta-dimers' phase with the hexagonal lattice metalens' phase profile. It is crucial to achieve coherent alignment, as the light scattered from both the meta-dimer and meta-atom arrays must constructively converge at the focal point (fig. S18).

In Fig. 5C, the hatched area in the diffraction diagram represents the spectrum where meta-dimers are integrated, which includes all overlap spectral regions where aliasing occurs. In this regard, the replacement of a high NA part with a meta-dimer array can significantly reduce the distorted diffraction. As a proof of concept, we fabricate two conventional metalens based on a square lattice and hexagonal lattice and anti-aliased metalens based on the integrated lattice (see Materials and Methods and fig. S19). Figure 5, E and F illustrate the measured intensity profiles at the focal plane for both integrated lattice metalens and conventional metalens (Optical setup: fig. S20). The aliasing radiating outward from the optical axis is substantially reduced in the integrated lattice metalens (fig. S21 and Movie 2). The measured (simulated) focusing efficiency of the integrated lattice metalens is around 39.2% (49.1%), while the focusing efficiencies of the hexagonal and square lattice metalenses are significantly lower, at approximately 20.3% (25.3%) and 18.2% (22.5%) respectively. The focusing efficiency is measured based on the input flux of the area where the aliased diffractions are not counted (fig. S22). Moreover, the measured peak intensity at the focal plane for the integrated lattice metalens is about twice as high as that of other hexagonal and square metalenses (fig. S23).

**Discussion**

In this study, we analyze aliasing in metasurfaces and develop effective anti-aliasing strategies. We discover that inadequate sampling leads to the outcoupling of higher-order spectra into the free space, manifesting as aliasing radiating in distorted directions. We find that the commonly employed Nyquist sampling criterion cannot prevent unintended diffraction. Moreover, we suggest that achieving aliasing-free reconstruction by a metasurface is not solely determined by its sampling periodicity; It requires a comprehensive understanding of the geometric interplay between the spectrum morphology and the sampling lattice structure. This insight leads to the development of anti-aliasing strategies for metasurfaces. For beam steering metasurfaces with directional spectra, rotation of the sampling lattice effectively eliminates aliasing by disrupting the coupling between the spectrum and higher-order light cones. In metalenses represented by isotropic spectra, the selection of the proper sampling lattice or coherent integration of optimized scatterers can significantly reduce aliasing phenomena. Anti-aliasing can not only eliminate undesired diffractions but also allow for larger unit cell sizes, which enables the complex design of meta-atoms (*32,40,41*) or relaxation of fabrication constraints. Additionally, the diffraction diagram obtained from light cone arrays helps identify the most effective diffractive direction for a given sampling lattice, in terms of both aliasing and efficiency. Therefore, based on the morphology of the desired spectrum, we can select a more effective sampling lattice. Additionally, exploring other sampling lattices, such as periodic Bravais lattices, quasi-periodic lattices, or even random lattices, which are not covered in this study, can be a very intriguing topic.

In conclusion, the aliasing in metasurface has been largely overlooked during rapid advancements in the field of metasurface. It is noted that aliasing should be regarded as a distinct issue, separated from the efficiency reduction of the desired diffraction. Aliasing that radiates unintended directions can serve as a non-negligible noise in optical devices. We emphasize that our findings are not limited to steering or focusing; they can be applied to the whole spectrum of metasurfaces, including holograms, structured beams, and other multi-functional wavefronts. With a deeper understanding of metasurface, we envision that our research paves the way for advanced high NA operation, particularly in the UV spectrum.

**References and Notes**

1. D. Lin, P. Fan, E. Hasman, M. L. Brongersma, Dielectric gradient metasurface optical elements. *Science* **345**, 298-302 (2014).
2. X. Yin, Z. Ye, J. Rho, Y. Wang, X. Zhang, Photonic Spin Hall Effect at Metasurfaces. *Science* **339**, 1405-1407 (2013).
3. L. Huang *et al.*, Three-dimensional optical holography using a plasmonic metasurface. *Nat. Commun.* **4**, 2808 (2013).
4. M. Khorasaninejad *et al.*, Metalenses at visible wavelengths: Diffraction-limited focusing and subwavelength resolution imaging. *Science* **352**, 1190-1194 (2016).
5. Y. Zhou, H. Zheng, I. I. Kravchenko, J. Valentine, Flat optics for image differentiation. *Nat. Photonics* **14**, 316-323 (2020).
6. J. Kim *et al.*, Scalable manufacturing of high-index atomic layer–polymer hybrid metasurfaces for metaphotonics in the visible. *Nat. Mater.* **22**, 474-481 (2023).
7. X. Fang, H. Ren, M. Gu, Orbital angular momentum holography for high-security encryption. *Nat. Photonics* **14**, 102-108 (2020).
8. Q. Song, M. Odeh, J. Zúñiga-Pérez, B. Kanté, P. Genevet, Plasmonic topological metasurface by encircling an exceptional point. *Science* **373**, 1133-1137 (2021).
9. J. Kim *et al.*, Photonic Encryption Platform via Dual-Band Vectorial Metaholograms in the Ultraviolet and Visible. *ACS Nano* **16**, 3546-3553 (2022).
10. J. Kim *et al.*, One-step printable platform for high-efficiency metasurfaces down to the deep-ultraviolet region. *Light Sci. Appl.* **12**, 68 (2023).
11. S. So *et al.*, Multicolor and 3D Holography Generated by Inverse-Designed Single-Cell Metasurfaces. *Adv. Mater.* **35**, 2208520 (2023).
12. J. Kim *et al.*, Dynamic Hyperspectral Holography Enabled by Inverse-Designed Metasurfaces with Oblique Helicoidal Cholesterics. *Adv. Mater.* **n/a**, 2311785.
13. I. Kim *et al.*, Metasurfaces-Driven Hyperspectral Imaging via Multiplexed Plasmonic Resonance Energy Transfer. *Adv. Mater.* **35**, 2300229 (2023).
14. A. Tittl *et al.*, Imaging-based molecular barcoding with pixelated dielectric metasurfaces. *Science* **360**, 1105-1109 (2018).
15. F. Yesilkoy *et al.*, Ultrasensitive hyperspectral imaging and biodetection enabled by dielectric metasurfaces. *Nat. Photonics* **13**, 390-396 (2019).
16. B. Wang *et al.*, Generating optical vortex beams by momentum-space polarization vortices centred at bound states in the continuum. *Nat. Photonics* **14**, 623-628 (2020).
17. M. Q. Mehmood *et al.*, Visible-Frequency Metasurface for Structuring and Spatially Multiplexing Optical Vortices. *Adv. Mater.* **28**, 2533-2539 (2016).
18. R. Juliano Martins *et al.*, Metasurface-enhanced light detection and ranging technology. *Nat. Commun.* **13**, 5724 (2022).
19. G. Kim *et al.*, Metasurface-driven full-space structured light for three-dimensional imaging. *Nat. Commun.* **13**, 5920 (2022).
20. G. J. Swanson, Binary optics technology: theoretical limits on the diffraction efficiency of multilevel diffractive optical elements.  (1991).
21. H. Chung, O. D. Miller, High-NA achromatic metalenses by inverse design. *Opt. Express* **28**, 6945-6965 (2020).
22. R. Menon, B. Sensale-Rodriguez, Inconsistencies of metalens performance and comparison with conventional diffractive optics. *Nat. Photonics* **17**, 923-924 (2023).
23. S. M. Kamali, E. Arbabi, A. Arbabi, A. Faraon, A review of dielectric optical metasurfaces for wavefront control. *Nanophotonics* **7**, 1041-1068 (2018).

24. J. Park *et al.*, All-solid-state spatial light modulator with independent phase and amplitude control for three-dimensional LiDAR applications. *Nat. Nanotechnol.* **16**, 69-76 (2021).
25. G.-Y. Lee *et al.*, Metasurface eyepiece for augmented reality. *Nat. Commun.* **9**, 4562 (2018).
26. Z. Li *et al.*, Inverse design enables large-scale high-performance meta-optics reshaping virtual reality. *Nat. Commun.* **13**, 2409 (2022).
27. R. Paniagua-Domínguez *et al.*, A Metalens with a Near-Unity Numerical Aperture. *Nano Lett.* **18**, 2124-2132 (2018).
28. H. Liang *et al.*, Ultrahigh Numerical Aperture Metalens at Visible Wavelengths. *Nano Lett.* **18**, 4460-4466 (2018).
29. M. Khorasaninejad, F. Capasso, Metalenses: Versatile multifunctional photonic components. *Science* **358**, eaam8100 (2017).
30. A. Ndao *et al.*, Octave bandwidth photonic fishnet-achromatic-metalens. *Nat. Commun.* **11**, 3205 (2020).
31. T. Shi *et al.*, Planar chiral metasurfaces with maximal and tunable chiroptical response driven by bound states in the continuum. *Nat. Commun.* **13**, 4111 (2022).
32. W. T. Chen *et al.*, A broadband achromatic metalens for focusing and imaging in the visible. *Nat. Nanotechnol.* **13**, 220-226 (2018).
33. Y. Yang *et al.*, Revisiting Optical Material Platforms for Efficient Linear and Nonlinear Dielectric Metasurfaces in the Ultraviolet, Visible, and Infrared. *ACS Photonics* **10**, 307-321 (2023).
34. D. Purves, J. A. Paydarfar, T. J. Andrews, The wagon wheel illusion in movies and reality. *Proc. Natl. Acad. Sci. U. S. A.* **93**, 3693-3697 (1996).
35. D. P. Petersen, D. Middleton, Sampling and reconstruction of wave-number-limited functions in N-dimensional euclidean spaces. *Information and Control* **5**, 279-323 (1962).
36. G. Zheng *et al.*, Metasurface holograms reaching 80% efficiency. *Nat. Nanotechnol.* **10**, 308-312 (2015).
37. A. Patri, S. Kéna-Cohen, C. Caloz, Large-Angle, Broadband, and Multifunctional Directive Waveguide Scatterer Gratings. *ACS Photonics* **6**, 3298-3305 (2019).
38. E. Khaidarov *et al.*, Asymmetric Nanoantennas for Ultrahigh Angle Broadband Visible Light Bending. *Nano Lett.* **17**, 6267-6272 (2017).
39. D. Sell, J. Yang, S. Doshay, R. Yang, J. A. Fan, Large-Angle, Multifunctional Metagratings Based on Freeform Multimode Geometries. *Nano Lett.* **17**, 3752-3757 (2017).
40. Z. Liu *et al.*, High-$Q$ Quasibound States in the Continuum for Nonlinear Metasurfaces. *Phys. Rev. Lett.* **123**, 253901 (2019).
41. S. An *et al.*, Multifunctional Metasurface Design with a Generative Adversarial Network. *Adv. Opt. Mater.* **9**, 2001433 (2021).
42. A. Arbabi *et al.*, Increasing efficiency of high numerical aperture metasurfaces using the grating averaging technique. *Sci. Rep.* **10**, 7124 (2020).
43. Swanson, G. J. Binary optics technology: the theory and design of multi-level diffractive optical elements. Vol. 854 (Massachusetts Institute of Technology, Lincoln Laboratory Cambridge, MA, USA, 1989).
44. Gupta, M. C. & Ballato, J. The handbook of photonics. (CRC press, 2018).
45. J. C. Heurtley, Scalar Rayleigh–Sommerfeld and Kirchhoff diffraction integrals: A comparison of exact evaluations for axial points*. *J. Opt. Soc. Am.* **63**, 1003-1008 (1973).

46. E. T. F. Rogers *et al.*, A super-oscillatory lens optical microscope for subwavelength imaging. *Nat. Mater.* **11**, 432-435 (2012).

## Methods

*Metasurface fabrication*

Beam steering metasurface, visible metalens, and UV metalens were fabricated on a 500-μm-thick silica substrate.

For the beam steering metasurface and visible metalens, an 800-nm-thick a-Si:H was deposited using plasma enhanced chemical vapor deposition (PECVD, BMR Technology HiDep-SC) with a flow rate of 10 sccm for SiH$_4$ and 75 sccm for H$_2$. Chamber pressure and operating temperature were 20 mTorr and 200 °C, respectively. Metasurfaces were transferred onto the positive photoresist (Microchem, 495 PMMA A2) by using a standard electron beam lithography process (ELINOIX, ELS-7800, acceleration voltage: 80 kV, beam current: 100 pA). Then, a 50-nm-thick chromium layer was deposited using electron beam evaporation (KVT, KVE-ENS4004). A lift-off process was proceeded and residual Cr patterns were used as a hard mask for the a-Si:H etching. After a dry etching process (DMS, silicon-metal hybrid etcher), the remaining chromium was removed using chromium etchant (CR-7).

For the UV metalens, a 450-nm-thick silicon nitride was deposited using plasma enhanced chemical vapor deposition (PECVD, BMR Technology HiDep-SC) with a flow rate of 35 sccm for SiH$_4$ and 35 sccm for H$_2$. Chamber pressure and operating temperature were 60 mTorr and 300 °C, respectively. Metasurfaces were transferred onto the positive photoresist (ZEON, ZEP520A) by using a high-speed electron beam lithography process (ELINOIX, ELS-BODEN 50H, acceleration voltage: 50 kV, beam current: 1 nA). Then, 50-nm-thick aluminum layer was deposited using electron beam evaporation (KVT, KVE-ENS4004). A lift-off process was proceeded and residual Al patterns were used as a hard mask for the silicon nitride etching. After a dry etching process (DRM85DD, TEL), the remaining chromium was removed using chromium etchant (CR-7).

*Calculating the diffraction efficiency of metasurface with the rigorous coupled-wave analysis*

We calculated the diffraction efficiency of a metasurface by the rigorous coupled-wave analysis (RCWA) method. Typically, metasurfaces are aperiodic, rendering the RCWA method unsuitable. However, if the number of samplings ($N$) within a projected period is a rational number, *i.e.* $N \times a = b$ for $a, b \in \mathbb{Z}$, the meta-atom array becomes periodic for the direction parallel to the phase gradient with the periodicity of $b\Lambda$. Here, $N$ can be represented as $\frac{\lambda}{\sin\theta \cdot \Lambda}$ where $\lambda$ is the wavelength of the incident light, $\theta$ is the deflection angle, and $\Lambda$ is the period of meta-atom. $\lambda/\sin(\theta)$ can be understood as the projected period so that the imparted momentums by ideal and aliasing order are $k_{\text{ideal}} = \frac{2\pi}{N\Lambda} = \frac{2\pi a}{b\Lambda}$ and $k_{\text{aliasing}} = k_{\text{ideal}} - G = \frac{2\pi(a-b)}{b\Lambda}$, respectively. The ideal and aliasing order of the metasurface can be defined as $a^{\text{th}}$ and $(a-b)^{\text{th}}$ diffraction order in the meta-atom supercell. For example, suppose a metasurface with the number of samplings as $N = 1.75$ for which the integers are $a = 4$ and $b = 7$. The ideal and aliasing order correspond to the 4$^{\text{th}}$ and 3$^{\text{rd}}$ diffraction order for respectively where the periodicity for supercell is 7Λ. Additionally, to account for the various meta-atom combinations within the super-periodicity, we calculate a set of diffraction efficiencies by varying the phase shift added to the phase profile and take the averaged efficiency as a representative value[42].

*Calculating the beam steering efficiency of metasurface*

In our study, the simulation of the metalens was conducted utilizing the three-dimensional Finite Difference Time Domain (FDTD) method, as implemented by Lumerical Inc. To ensure accuracy and minimize reflections, Perfectly Matched Layer (PML) boundary conditions were applied along all three axes. Our approach involved a full-wave simulation of the metalens, distinctively omitting the utilization of far-field propagation techniques. The diameter of metalens used for calculation is 25 μm, considering the computational cost.

**Data availability**

The data that support the findings of this study are available from the corresponding author upon reasonable request.

**Author contributions**

J.R. and S.K. conceived the idea and initiated the project; S.K. and J.K. designed the experiments; S.K. and K.K. performed the theoretical and numerical simulations; J.K. fabricated the nanopatterns and devices; S.K., J.K., K.K. and M.J. performed the experimental characterization and data analysis; S.K. mainly wrote the paper; All authors contributed to the discussion and analysis and have given approval of the final version of the paper; J.R. guided the entire project.

**Acknowledgements**

This work was financially supported by the POSCO-POSTECH-RIST Convergence Research Center program funded by POSCO, the National Research Foundation (NRF) grants (NRF-2022M3C1A3081312, NRF-2022M3H4A1A02074314, NRF-2019R1A5A8080290, RS-2023-00302586, RS-2023-00283667) funded by the Ministry of Science and ICT of the Korean government, and the Korea Evaluation Institute of Industrial Technology (KEIT) grant (No. 1415179744/20019169, Alchemist project) funded by the Ministry of Trade, Industry and Energy (MOTIE) of the Korean government. J.K. acknowledges the Asan Foundation Biomedical Science fellowship and the Presidential Science fellowship funded by the Ministry of Science and ICT (MSIT) of the Korean government.

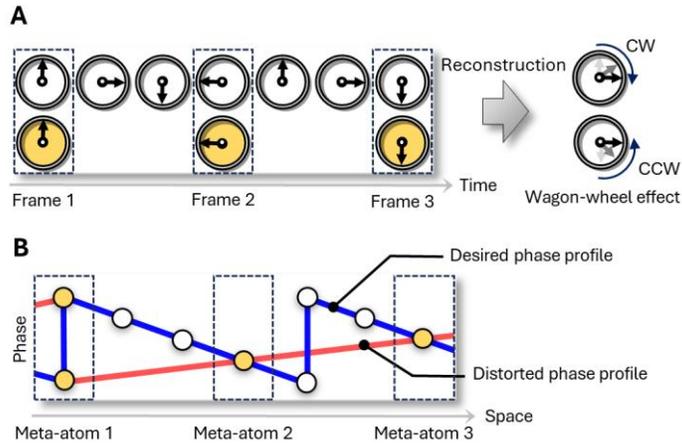

**Fig. 1. The wagon-wheel effect in metasurface.** (**A**) Schematic illustration of the temporal sampling system. The arrow rotating in a clockwise (CW) direction is captured over a few frames in time, leading observers to perceive it as rotating counterclockwise (CCW) after reconstruction. This visual illusion is commonly referred to as the 'wagon-wheel effect' (*34*). (**B**) Schematic illustration of spatial sampling in a metasurface. The desired phase profile (blue line) is sampled by a few meta-atoms on the metasurface. However, the set of sampled phases (denoted as yellow circles) also can reconstruct another phase profile (red line) which has an opposite sign of slope. This ambiguity leads to the spatial wagon-wheel effect occurring in the metasurface, manifesting as unwanted diffractions having distorted transverse momenta.

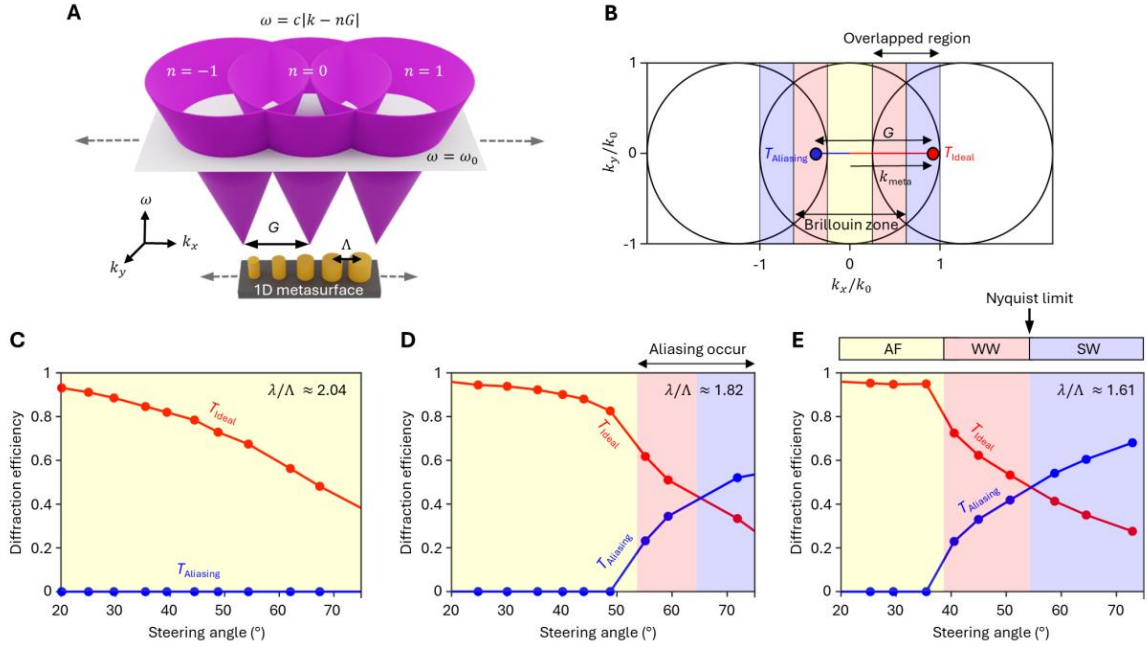

**Fig. 2. Aliasing in 1D beam steering metasurface.** (**A**) Schematic illustration of 1D metasurface and reproduced light cones in the momentum space. The spectral range where aliasing occurs can be obtained from overlapping light cones. Depending on the target frequency of light ($\omega_0$) and the sampling interval ($\Lambda$), the range of overlap spectral region is determined. (**B**) The cross-section of light cones at the target frequency. Depending on the magnitude of the encoded momentum ($|k_{meta}|$), the normally incident light can couple not only the $0^{th}$-order light cone but also the adjacent higher-order light cone. (**C-E**) Calculated steering efficiency for various sampling levels where (C) $\lambda/\Lambda = 2.04$, (D) $\lambda/\Lambda = 1.82$, (E) $\lambda/\Lambda = 1.61$, respectively. Yellow, red, and blue colors in the graphs represent the three diffraction regimes, aliasing-free (AF), weak wagon-wheel effect (WW), and strong wagon-wheel effect (SW) regimes, respectively.

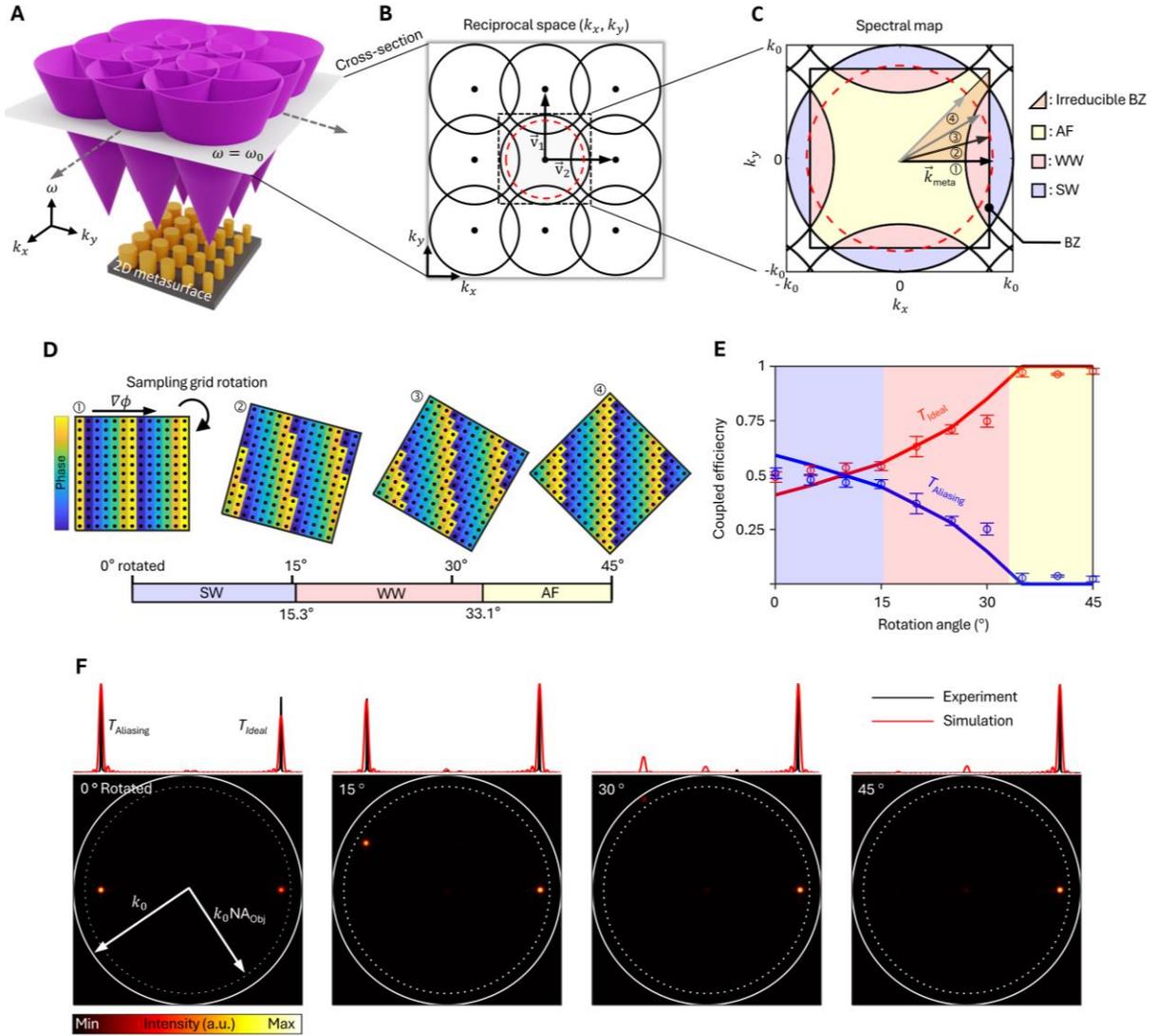

**Fig. 3. Anti-aliasing of metasurface by rotating sampling lattice.** (**A**) Schematics of a metasurface and its corresponding light cone array in the momentum space. (**B**) Cross-section of the light cones at the target frequency ($\omega = \omega_0$). (**C**) Diffraction diagram obtained from overlap light cones, identifying different diffraction regimes. The regimes of AF, WW, and SW are indicated in yellow, red, and blue, respectively. (**D**) Illustration of rotated sampling lattices with an identical phase map, highlighting how diffraction regimes change in terms of the rotation angle of the sampling lattice. (**E**) The efficiency ratio between ideal and aliasing order as a function of the sampling lattice's rotation angle. The measurements were repeated 5 times; error bars represent the standard deviation. The solid line represents calculated data from finite-difference time-domain (FDTD) simulation. (**F**) Presentation of experimental (black line) and calculated (red line) far-field radiation pattern (top). The momentum space images (bottom) are calculated from FDTD simulation. The outer and inner dotted circles in the momentum space image represent the light cone and the NA of the objective lens, respectively.

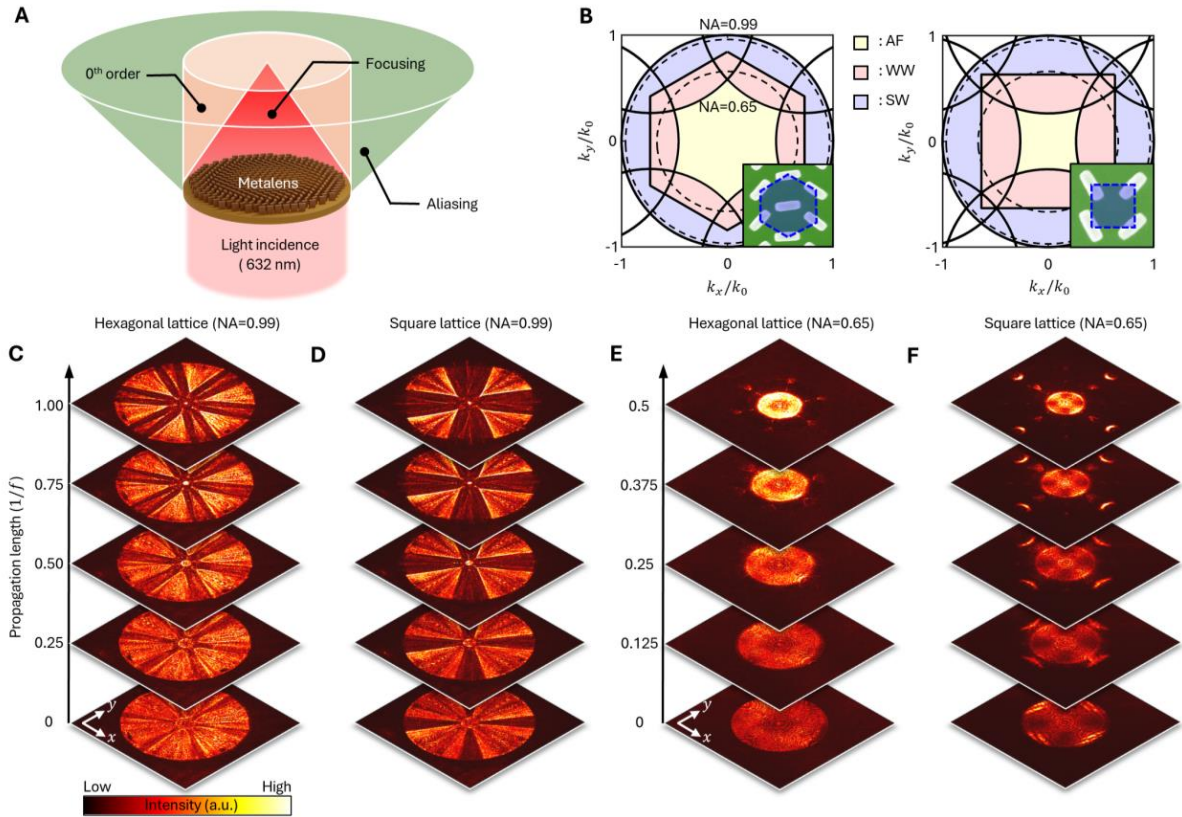

**Fig. 4. Aliasing in metalens.** (**A**) The three dominant diffraction channels in high NA metalens. (**B**) Diffraction diagrams in a hexagonal lattice (left) and square lattice (right) and scanning electron microscope (SEM) image of the fabricated metalens (insets). The inner and outer dotted circles represent the spectral boundary having NA 0.99 and 0.65. (**C, D**) The measured intensity distribution of diffracted light after transmitting (C) the hexagonal lattice and (D) square lattice metalens with an NA of 0.99. (**E, F**) The measured intensity distribution of diffracted light after transmitting (E) the hexagonal lattice and (F) square lattice metalens with an NA of 0.65.

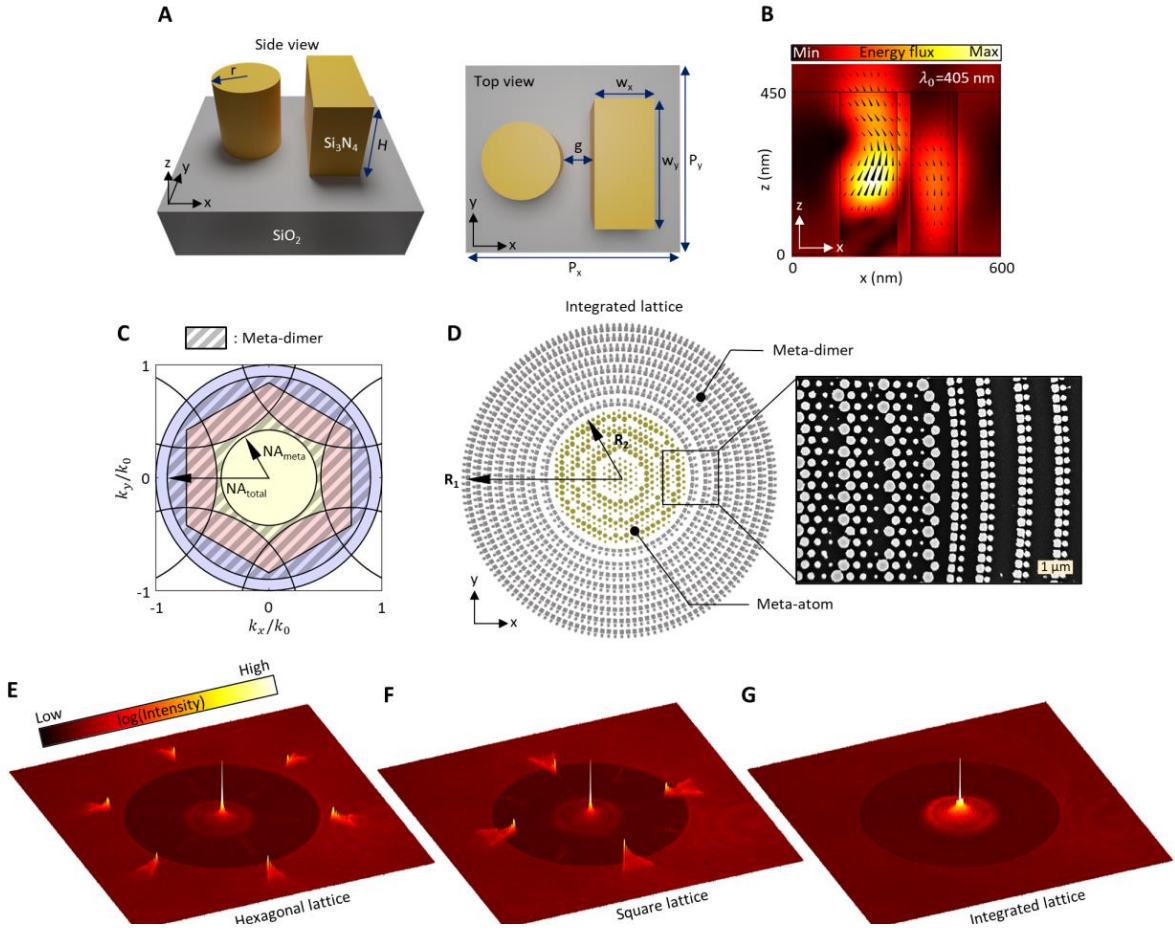

**Fig. 5. Anti-aliasing metalens via integration of meta-dimer.** (**A**) Schematic of meta-dimer consists of two coupled vertical waveguides. (**B**) Calculated time averaged Poynting vector in *xz*-plane of meta-dimer. (**C**) Diffraction diagram of hexagonal metalens. Hatched area represents the spectrum where meta-dimers are integrated, including all the spectrum where aliasing occurs. (**D**) Schematic illustration of meta-dimer integrated metalens and SEM image of the fabricated UV metalens. (**E-G**) Intensity distribution at the focal plane for (E) hexagonal lattice metalens, (F) square lattice metalens, and (G) meta-dimer integrated metalens.

*Supplementary Materials for*
# Anti-aliased metasurfaces beyond the Nyquist limit

Seokwoo Kim[1]†, Joohoon Kim[1]†, Kyungtae Kim[1]†, Minsu Jeong[1], and Junsuk Rho[1,2,3,4,5]*

[1]Department of Mechanical Engineering, Pohang University of Science and Technology (POSTECH), Pohang 37673, Republic of Korea

[2]Department of Chemical Engineering, Pohang University of Science and Technology (POSTECH), Pohang 37673, Republic of Korea

[3]Department of Electrical Engineering, Pohang University of Science and Technology (POSTECH), Pohang 37673, Republic of Korea

[4]POSCO-POSTECH-RIST Convergence Research Center for Flat Optics and Metaphotonics, Pohang 37673, Republic of Korea

[5]National Institute of Nanomaterials Technology (NINT), Pohang 37673, Republic of Korea

*Corresponding author. E-mail: jsrho@postech.ac.kr

## Supplementary Text

Aliasing in 1D phase gradient metasurface.

Here, we provide an analysis of diffraction in a one-dimensional (1D) phase gradient metasurface where a 1D sampling grid is employed. Here, the meta-atoms are periodically arranged along the $x$-axis at a subwavelength interval of $\Lambda$ and sample the phase profile $\phi(x)$. Note that $\phi(x)$ always exhibits a band-limited spectrum restricted by the wavenumber since generally, the metasurface is targeted for propagating light along free space. Following the scalar approach (*43*), the complex amplitude encoded by meta-atoms can be approximated as,

$$t(x) = \sum_{n=-\infty}^{\infty} \delta(x - \Lambda n)\, e^{i\phi(x)} \otimes g(x),$$

where $\otimes$ represents a convolution and $g(x)$ represents the local transmittance obtained from an individual meta-atom. Without loss of generality, we assume that $g(x)$ is invariant for each meta-atom. The encoded spectrum on the metasurface can be obtained by taking the Fourier transform of $t(x)$, which is,

$$T_{\text{encoded}}(k) = \frac{w(k)}{\Lambda} \sum_{n=-\infty}^{\infty} F(k - nG),$$

where $n$ is an integer number and $G = \frac{2\pi}{\Lambda}$. Here, $F(k)$ is the desired spectrum from the phase profile, and the weighting function $w(k)$ is a Fourier transformation of local transmittance $g(x)$. One can find that the encoding procedure reproduces the spectrum periodically at an interval of $G$ (Fig. S1). Among them, the higher-order spectra ($F(k - nG)$ where $n \neq 0$) possesses shifted momentum by $nG$ compared to the desired spectrum $F(k)$.

To consider only the radiative field that can propagate into the free space, the light-cone filter, which filters out the evanescent field, must be multiplied by the encoded spectrum,

$$T_{\text{radiative}} = T_{\text{encoded}} \cdot H_{\text{LC}},$$

where piecewise function $H_{\text{LC}}$ is zero outside the interval $[-k_0, k_0]$ and unity inside it. One can find that the diffraction of phase gradient metasurface is then determined by three different functions: repetitive spectra $F(k - nG)$, the weighting function $w(k)$, and the light cone filter $H_{\text{LC}}$. First, the weighting function $w(k)$ has to do with the diffraction efficiency of each momentum. For example, if we assume the local transmittance function $g(x)$ for meta-atoms is a rectangular function whose interval is $\Lambda$, the diffraction efficiency of the radiative spectrum then becomes $|\text{sinc}(k_{\text{meta}}/G)|^2$ where $k_{\text{meta}}$ is an arbitrary momentum within the desired spectrum (*44*). The *sinc*-like shape of the weighting function $w(k)$ implies an important fact: as the encoded momentum $k_{\text{meta}}$ increases, then its diffraction efficiency decreases, and vice versa. Next, the light cone determines the spectral area where aliasing occurs. The momenta included in high-order spectra become aliasing of diffraction when the part of higher-order spectra is inside the light cone ($|k| < k_0$). Therefore, the emergence of aliasing is determined by the geometric relation between repetitive spectra and light cones (Fig. S1). Since the higher-order spectra have shifted momentum by $nG$, aliased spectra radiate in unwanted directions.

Constructing a diffraction diagram by repetitive light cones.

    In the encoded spectrum of a metasurface, as illustrated in Fig. S1, identifying the specific spectral regions subject to distortion during diffraction presents a challenge. A diffraction diagram, serving as a type of spectrum-based representation, presents the regions anticipated to undergo distorted diffraction. This diffraction diagram is derived by modifying the momentum in the reciprocal space ($k \to k + nG$ for $n \in \mathbb{Z}$), an approach we term the 'spectral point of view'. This method facilitates a clearer understanding of the spectral behavior within metasurfaces. The relationship between repetitive spectra $F(k - nG)$ and light cone filter $|k| < k_0$ is transformed into the interplay between repetitive light cones $|k + nG| < k_0$ and ideal spectrum $F(k)$. Thus, the aliasing is understood as the coupling between the ideal spectrum and higher-order light cones (for $n \neq 0$). The distorted momentum of aliasing can be restored from the previously used substitution relationship $k \to k + nG$. For example, the momentum $k$, coupled with the $n^{\text{th}}$-order light cone becomes $k - nG$ in free space. Once again, the main advantage of this spectral perspective is the identification of the spectral regions susceptible to aliasing. For instance, as illustrated in Fig. S2a and c, aliasing occurs if the spectrum resides within a region of overlapping light cones since it couples not only to the $0^{\text{th}}$-order light cone but also to the higher-order light cone. In instances where the Nyquist sampling criterion is violated, as shown in Fig. S1b and d, the spectrum is positioned outside the Brillouin zone boundary, indicating that it couples more deeply with higher-order light cones compared to the 0th-order light cone, which corresponds to the strong wagon-wheel (SW) regime.

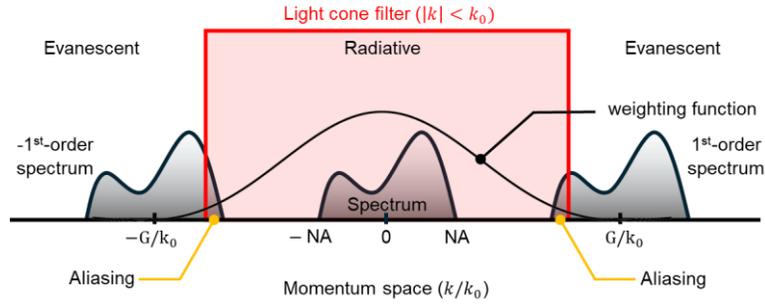

**Fig. S1. The occurrence of aliasing in 1D phase gradient metasurface.** The discrete sampling of the metasurface leads to the reproduction of the spectrum at an interval of $G = 2\pi/\Lambda$, where $\Lambda$ is the periodicity of meta-atoms. The weighting function, which is the Fourier transform of the local transmittance function for each meta-atom, contributes to the diffraction efficiency of each momentum in repetitive spectra. The momentum smaller than the wavevector ($k_0 = 2\pi/\lambda_0$ where $\lambda_0$ is the target wavelength) can propagate into free space. The radiative momenta from higher-order spectra become aliasing of the metasurface. Note that the spectrum is depicted symmetrically, but it is more common for it to be asymmetric.

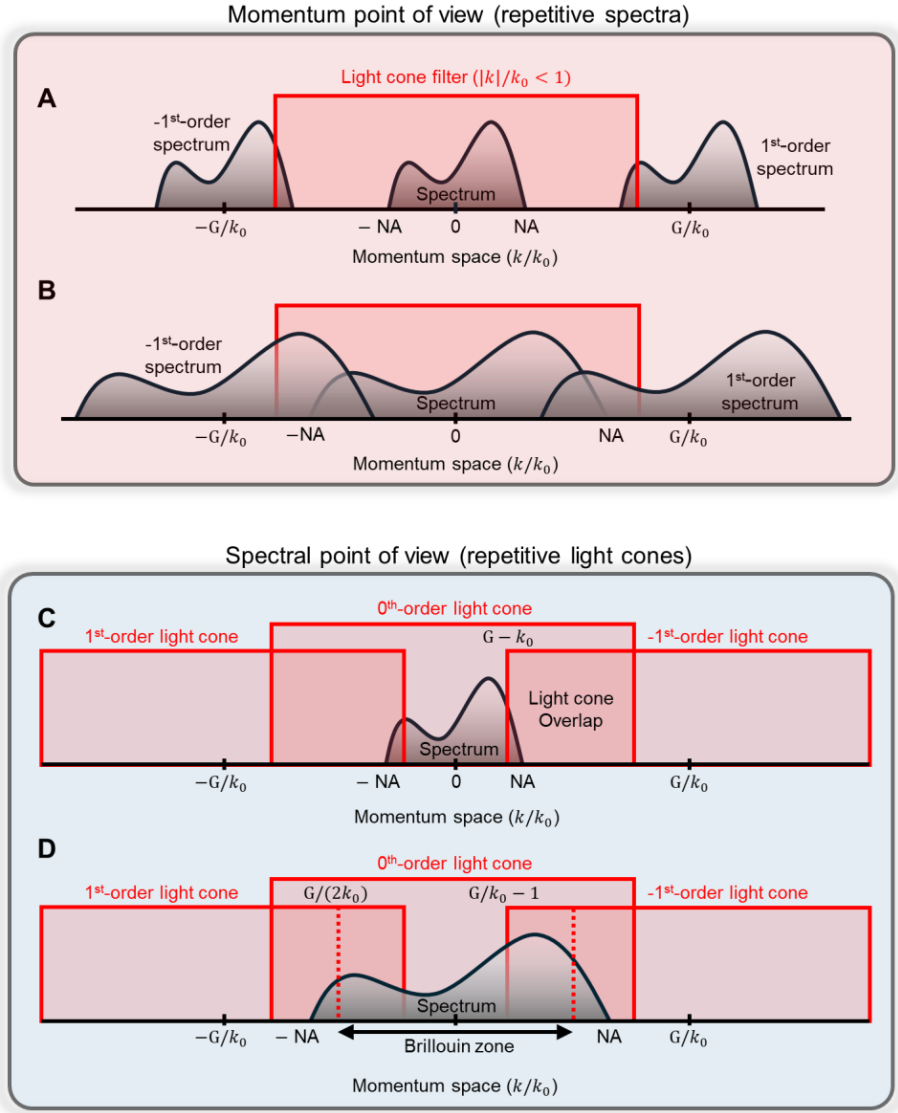

**Fig. S2. Constructing the diffraction diagram through light cone array.** (**A**) The schematic of the encoded spectrum of metasurface in the reciprocal space. Spectrum is a band-limited function, and its upper and lower bounds are assumed to be equal, *i.e.* $|k_{\text{meta}}/k_0| < \text{NA}$. The aliasing happens when higher-order spectra are located inside the light cone. (**B**) The overlap spectrum belongs to strong wagon-wheel regimes which doesn't satisfy the Nyquist criterion. (**C**) The schematic of the encoded spectrum of metasurface from a spectral point of view. The light cones are reproduced at an interval of *G*. The spectrum belongs to the overlap region between higher-order light cones to become aliasing channels. (**D**) The part of the spectrum located outside of the Brillouin zone belongs to strong wagon-wheel regimes which doesn't satisfy the Nyquist criterion.

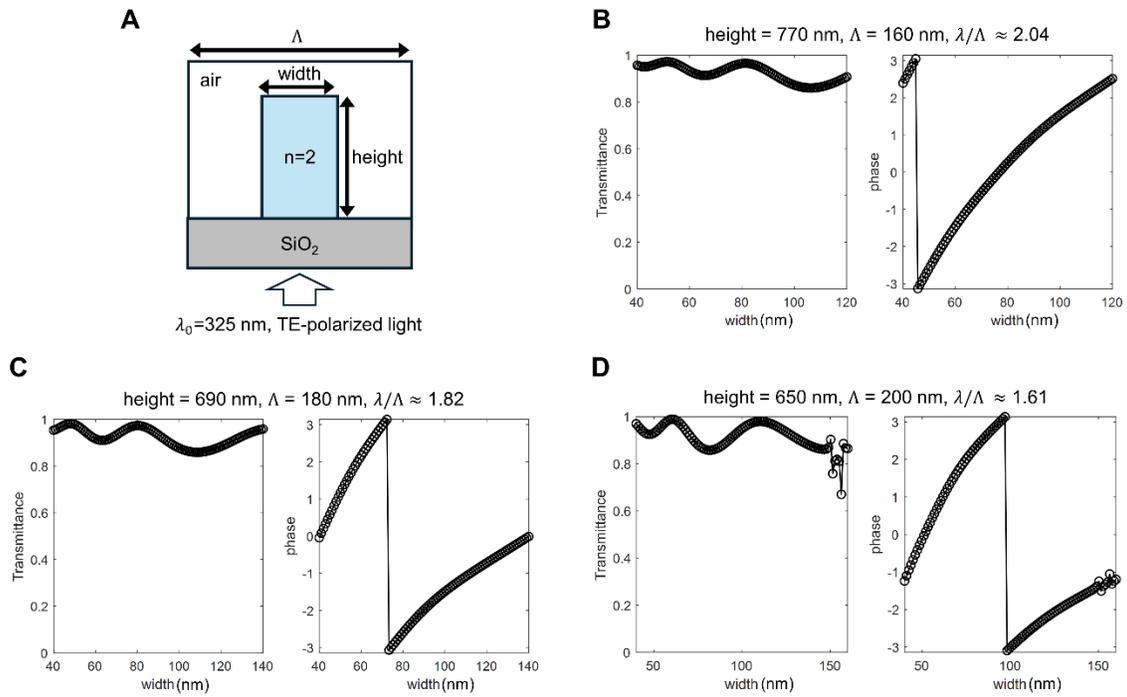

**Fig. S3. Meta-atom information for 1D beam steering metasurface.** (**A**) Schematic illustration of meta-atom used for Fig. 2C-E. (**B-D**) The transmittance and phase of meta-atom set used for calculating the beam steering efficiency in Fig. 2C, D, and E. All data are calculated using the RCWA.

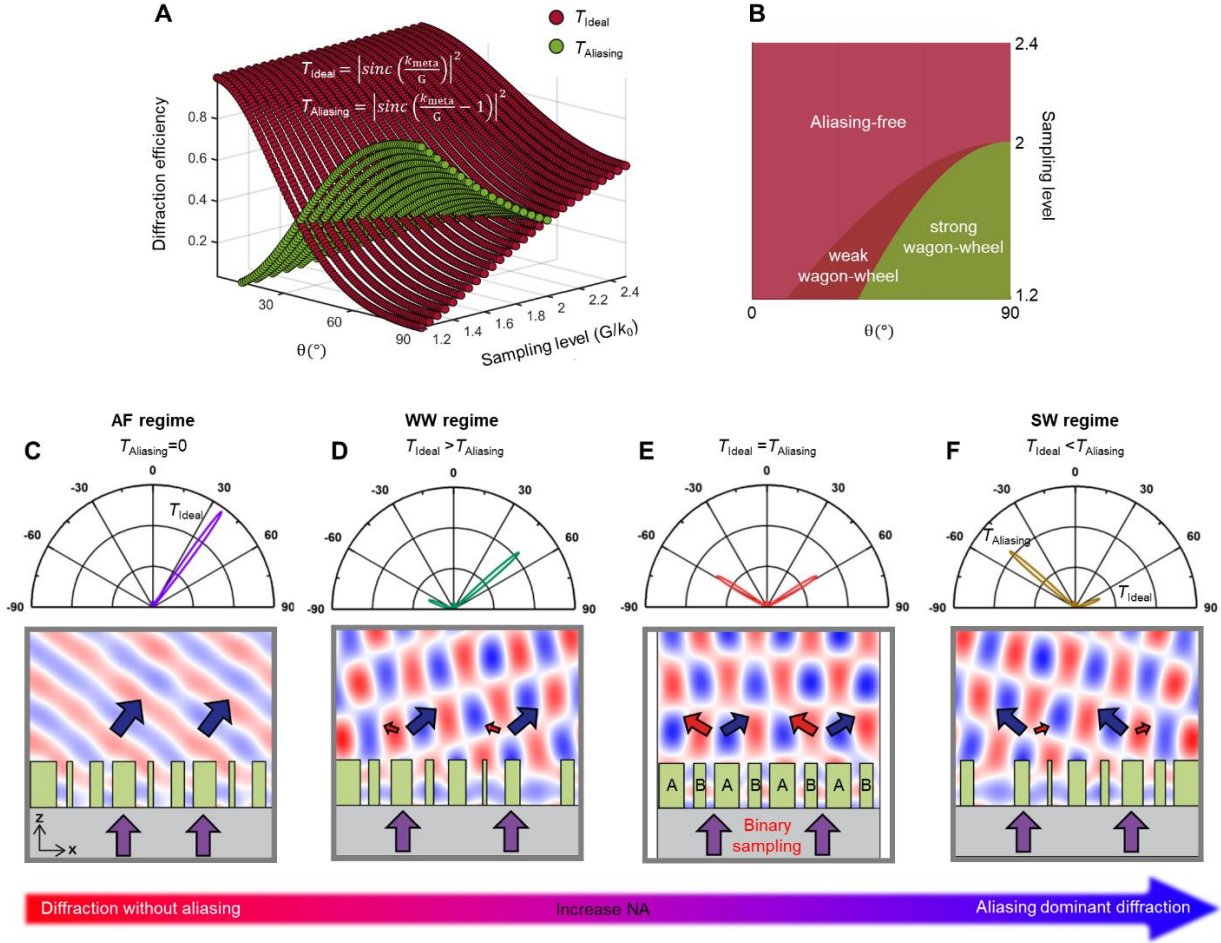

**Fig. S4. Aliasing in 1D beam steering metasurface.** (**A**) Calculated Diffraction efficiency of both ideal and aliasing diffraction orders using scalar diffraction approximation (*43*). Here, the local transmittance function of the meta-atom is assumed a rectangular function; therefore, the efficiency of ideal order is approximated as $|sinc(k_{meta}/G)|^2$. The aliasing order is the 1$^{st}$ or -1$^{st}$ diffraction order, which is diffracted by phase error, *i.e.* the phase difference between the desired phase profile and the sampled phase profile. The grating vector of phase error is *G*, therefore, the efficiency of aliasing order becomes $|sinc(k_{meta}/G - 1)|^2$ (for $k_{meta} > 0$). (**B**) The phase map of diffraction regimes in 2D parametric space. When the sampling level falls below 2, WW and SW regimes occur, with the range of these regimes expanding as the sampling level decreases. (**C-F**) The simulated radiation patterns of transmitted light from the beam steering metasurfaces and their electric field intensity distribution for four different steering angles. The steering angle for ideal and aliasing order is asymmetric except for the case whose diffraction regime is on the transition point from WW to SW (E). Note that the meta-atom array in (D) is inversion symmetric with the meta-atom array in (F).

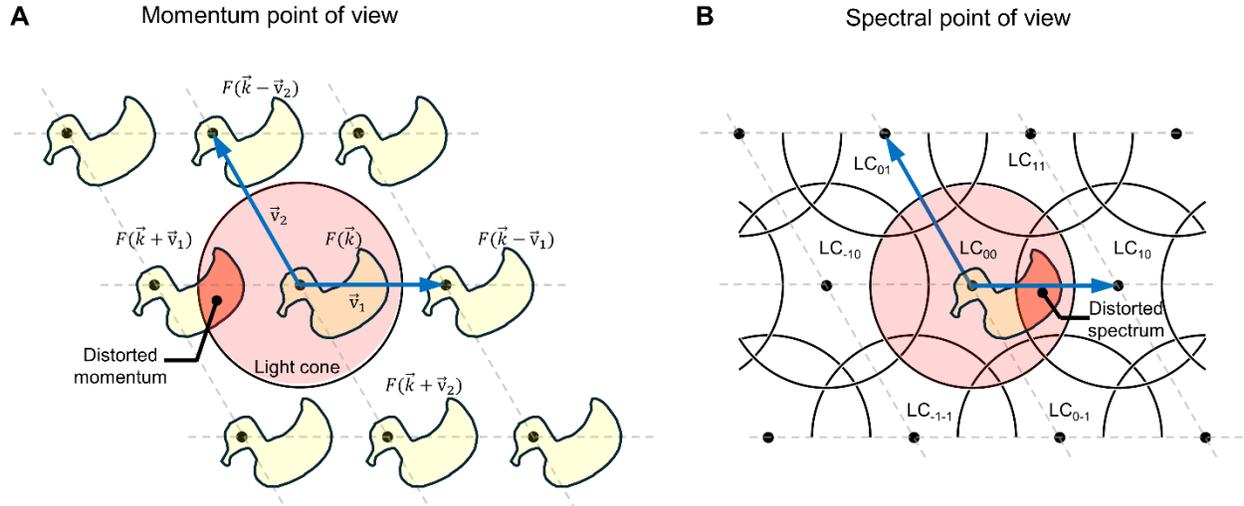

**Fig. S5. Constructing a 2D diffraction diagram.** (**A**) When the phase profile is sampled by 2D sampling lattice, its spectrum $F(\vec{k})$ (yellow-colored area) is replicated on the points generated by the linear combination of two nonparallel reciprocal vectors $\vec{v} = n\vec{v}_1 + m\vec{v}_2$ for integer $n$ and $m$. The light cone centered origin filters out the spectrum that cannot propagate to the free space, that is, the evanescent field. As shown, when higher-order spectra $F(\vec{k}-\vec{v})$, which having shifted momenta by $\vec{v}$, overlap with a light cone, the overlapping spectrum becomes an aliasing radiating with distorted momenta. (**B**) By momentum substitution $\vec{k} \to \vec{k}+\vec{v}$ (see Supplementary text section "Constructing a diffraction diagram by repetitive light cones"), the diffraction diagram (red-colored spectral region) can be obtained from a repetitive light cone array ($LC_{nm}$). A diffraction diagram can identify the spectral range where aliasing can occur. For instance, when the spectrum falls within the spectral region where light cones overlap, aliasing occurs as it couples with the higher-order light cone. The distorted momentum of the aliasing order can be restored by $\vec{k} \to \vec{k}-\vec{v}$.

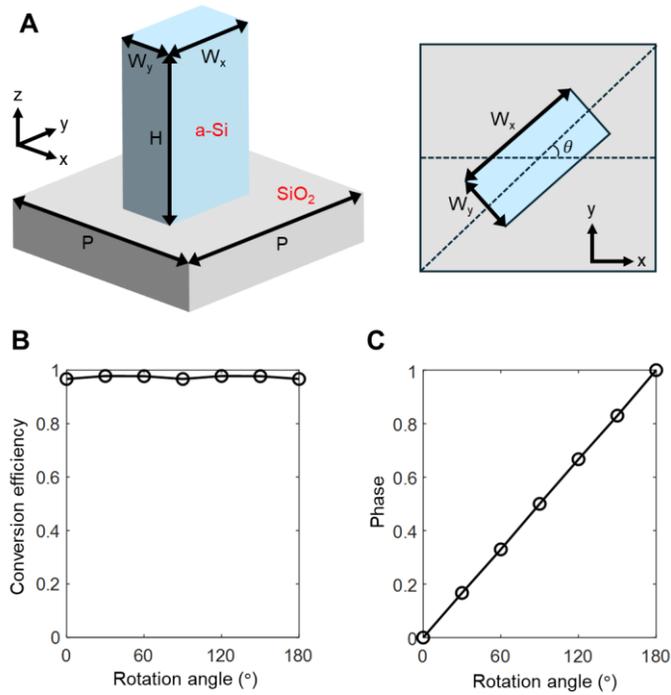

**Fig. S6. The design of meta-atom for beam steering metasurface in visible wavelength.** (**A**) Schematic representation of the unit cell: P = 400 nm, H = 700 nm, Wx = 320 nm, Wy = 95 nm. The refractive index of hydrogenated amorphous silicon (a-Si) is 2.42 at the target wavelength of 632 nm. (**B, C**) The graphs depict the conversion efficiency, and geometric phase as a function of the rotation angle.

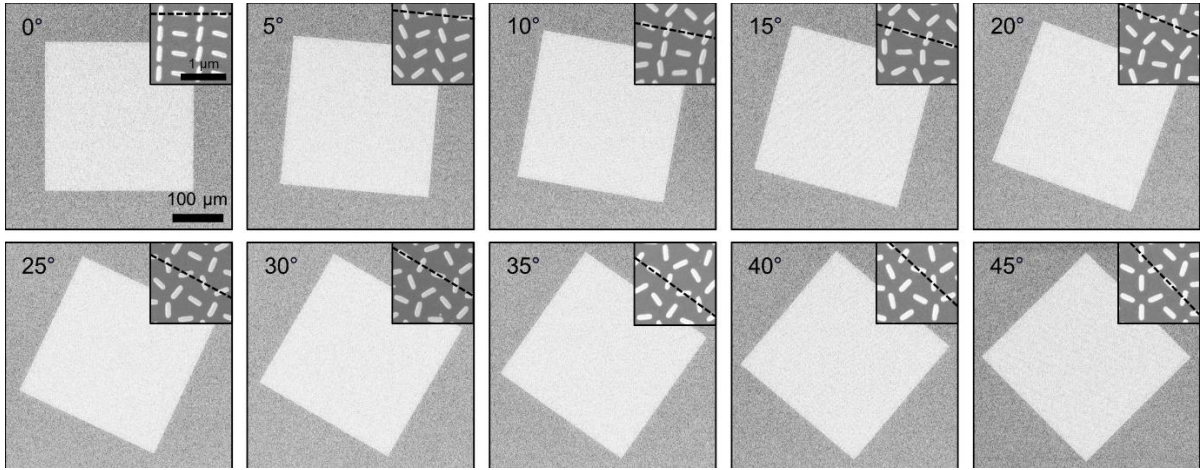

**Fig. S7. SEM images for rotating beam steering metasurfaces.** A total of ten beam steering metasurfaces are fabricated, with each metasurface subjected to a sampling grid rotation incremented at 5° intervals. The dimensions of each metasurface are 300 µm × 300 µm. A magnified SEM image of the meta-atoms for each metasurface has been included as an inset.

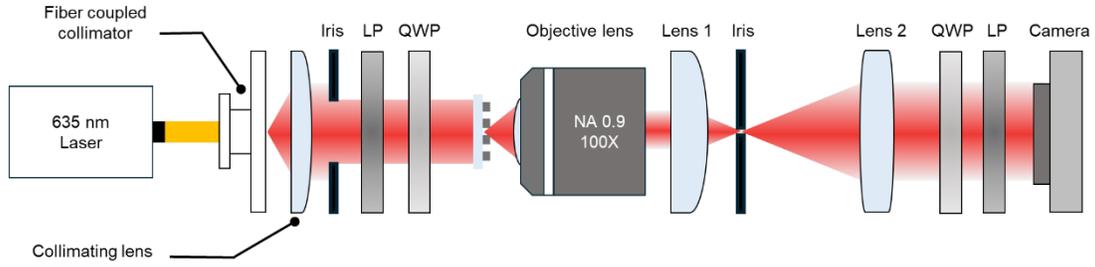

**Fig. S8. The schematic illustration of the optical setup for measuring the momentum space of the beam steering metasurface.** QWP: quarter-wave-plate, LP: linear polarizer. The laser beam is collimated by a single-mode fiber-coupled collimator and collimating lens. It becomes left circularly polarized after passing through the LP and QWP. An objective (100× magnification, NA = 0.9) is used to collect the laser beam deflected by the beam steering metasurface. We used a 4f lensing system with a lens 1 (f = 3 cm) and lens 2 (f = 9 cm) to relay the spatial Fourier transformation of a laser beam in the back focal plane to the entrance aperture coupled with the sCMOS camera. To filter out the $0^{th}$-order light, QWP and LP are placed after the 4f lensing system.

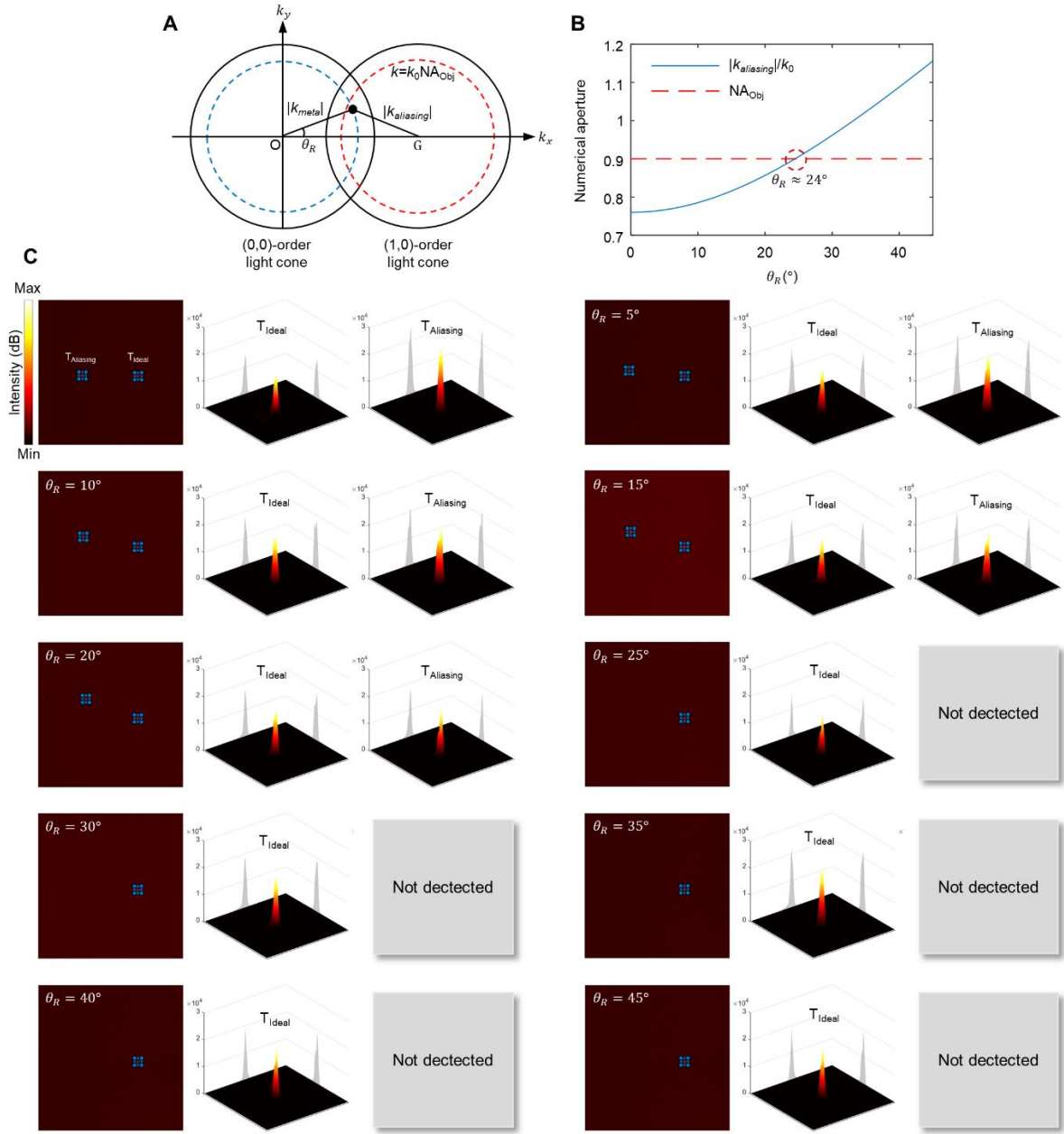

**Fig. S9. Measured momentum space images for beam steering metasurfaces.** (**A**) Schematic diagram representing the relationship between the momentum of aliasing ($k_{aliasing}$) and the rotation angle of the sampling grid ($\theta_R$). The blue and red dotted circles indicate the magnitudes of the $k_{meta}$ and NA of the objective lens used for the measurement, respectively. (**B**) The correlation of NA of aliasing order and $\theta_R$. Notably, at a $\theta_R$ of approximately 24°, the NA of the aliasing order surpasses that of the objective lens, rendering it non-measurable. (**C**) Measured data for momentum space imaging. Above $\theta_R = 25°$, the aliasing order is not captured on camera.

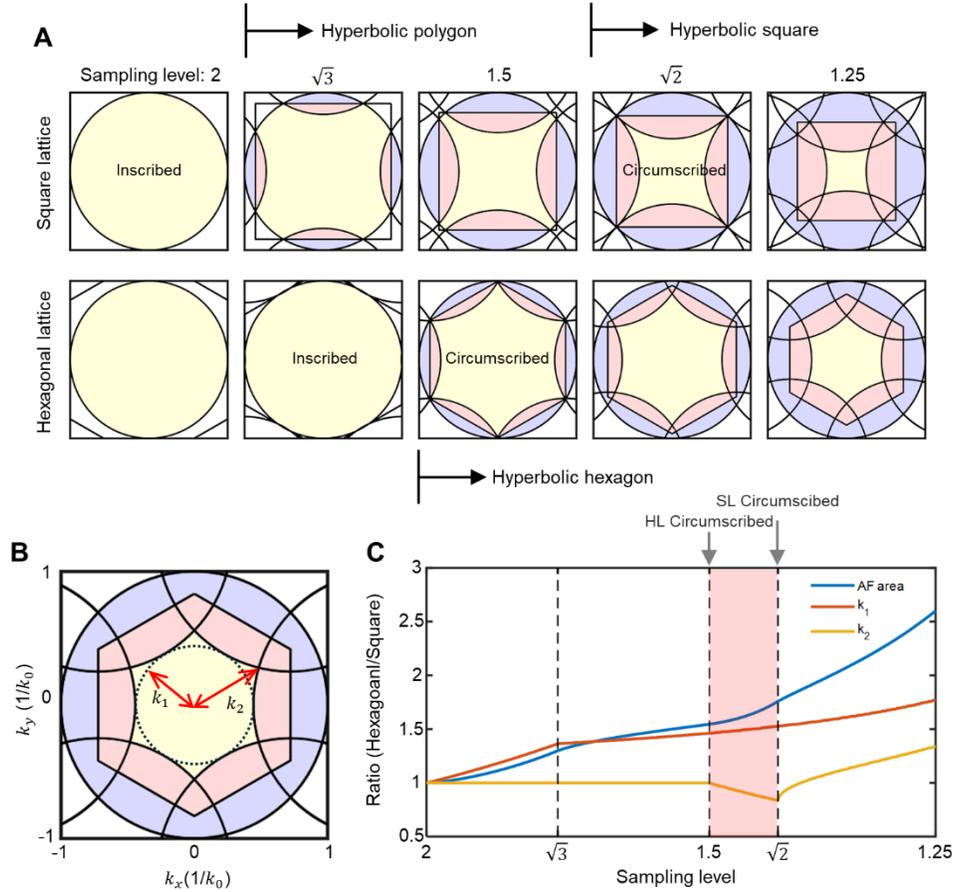

**Fig. S10. Comparative analysis of hexagonal and square sampling lattice in terms of aliasing.** To compare the two most renowned sampling lattices, square and hexagonal lattice, we analyze the diffraction diagrams for an arbitrary sampling level ($\lambda/\Lambda$). (**A**) The evolution of spectral morphology as sampling level decreases. Clearly, the morphology of the AF spectral region depends on its sampling level. Especially when the AF spectral region is inscribed and circumscribed by the Brillouin zone boundary, it undergoes a morphology transition, *e.g.* circle to hyperbolic polygon. (**B**) The parameters $k_1$ and $k_2$ represent the maximum NA for an isotropic spectrum and the maximum magnitude of momentum belongs to the AF region, respectively. (**C**) The ratio of three quantities, area of AF region, and $k_{1,2}$ between the hexagonal lattice to the square lattice from the sampling level from 2 to 1.25. In most cases, the hexagonal lattice is superior to the square lattice (Ratio > 1). Counterintuitively, for $k_2$, the square lattice surpasses the hexagonal lattice for sampling levels between 1.5 to $\sqrt{2}$. This is attributed to the morphology of two lattices; the circumscribed step for a hexagonal lattice occurs at more higher sampling level ($\lambda/\Lambda = 1.5$) than a square lattice ($\lambda/\Lambda = \sqrt{2}$), rendering the AF region to lie inside the light cone.

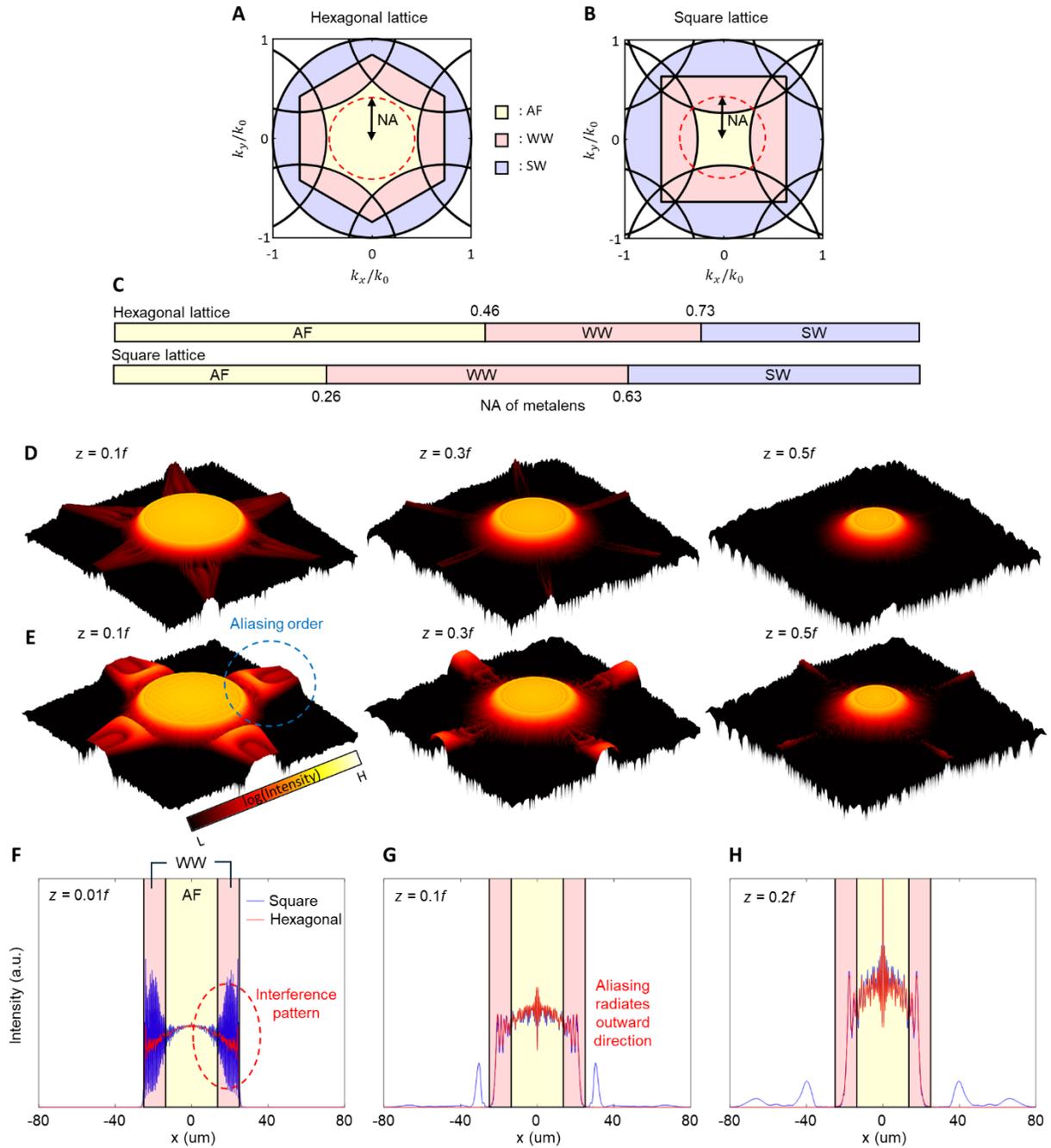

**Fig. S11. Anti-aliasing with selection of proper lattice.** (**A, B**) The diffraction diagrams for hexagonal lattice (left) and square lattice (right), respectively. The sampling level of metalenses is set as 1.264. The inscribed red dotted boundary represents the NA (0.45) of metalens. (**C**) The range of NA in terms of the diffraction regimes for hexagonal lattice (top) and square lattice (bottom). One can observe that hexagonal lattice has more relaxed conditions than square lattice in that the hexagonal lattice has more wider AF range. (**D, E**) The evolutions of intensity distribution diffracted from the sampled hyperbolic lens phase profile calculated from the scalar diffraction theory (*45*). NA and the sampling level of metalenses are set as 0.45 and 1.264, respectively. The incident light is diffracted from the 2D complex amplitude array whose phase data correspond to the sampled hyperbolic lens phase profile and their magnitudes are assumed

as unity. (**F-H**) The *x*-axis cross-section of intensity profiles for square lattice (blue line) and hexagonal lattice (red line) at different propagation lengths. The diameter of metalens is set as 40 µm. The yellow and red colored area within the metalens ($|x| < 20$ um) corresponds to the AF and WW diffraction regimes for square lattice, obtained from its diffraction diagram. The fluctuated pattern shown in (F) is attributed to the interferences between ideal order and aliasing order, occurring in WW regimes. The aliasing order radiates outward direction as propagation length increases.

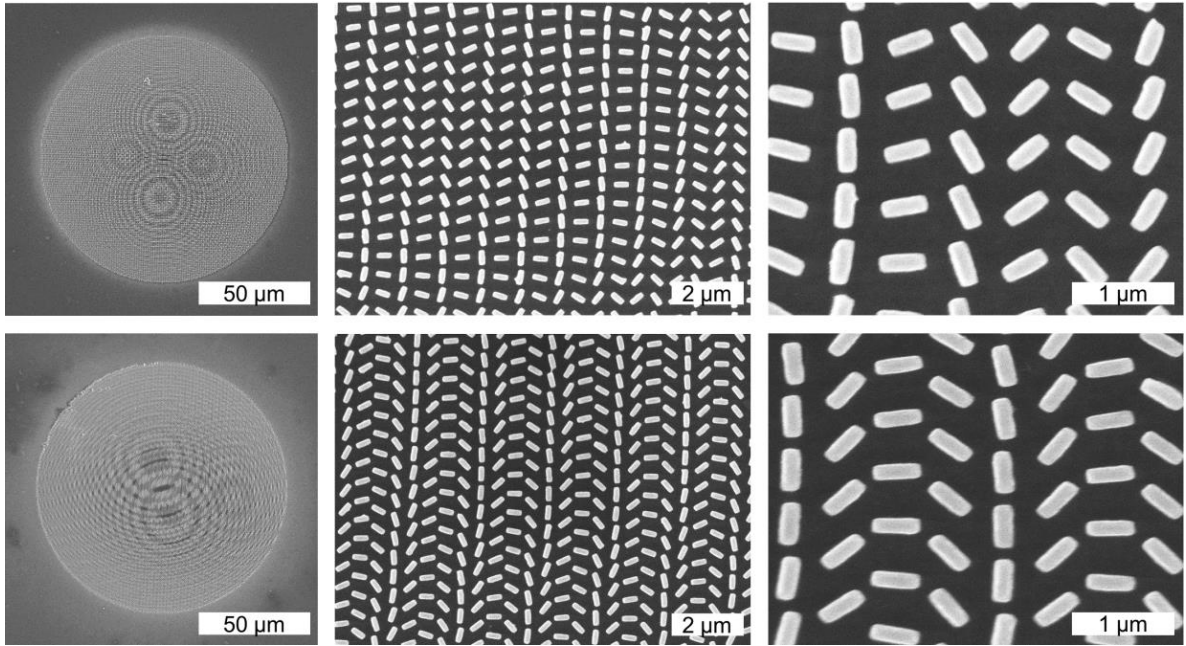

**Fig. S12. SEM images of fabricated visible metalenses.** The SEM images of fabricated metalenses based on square lattice (top) and hexagonal lattice (bottom).

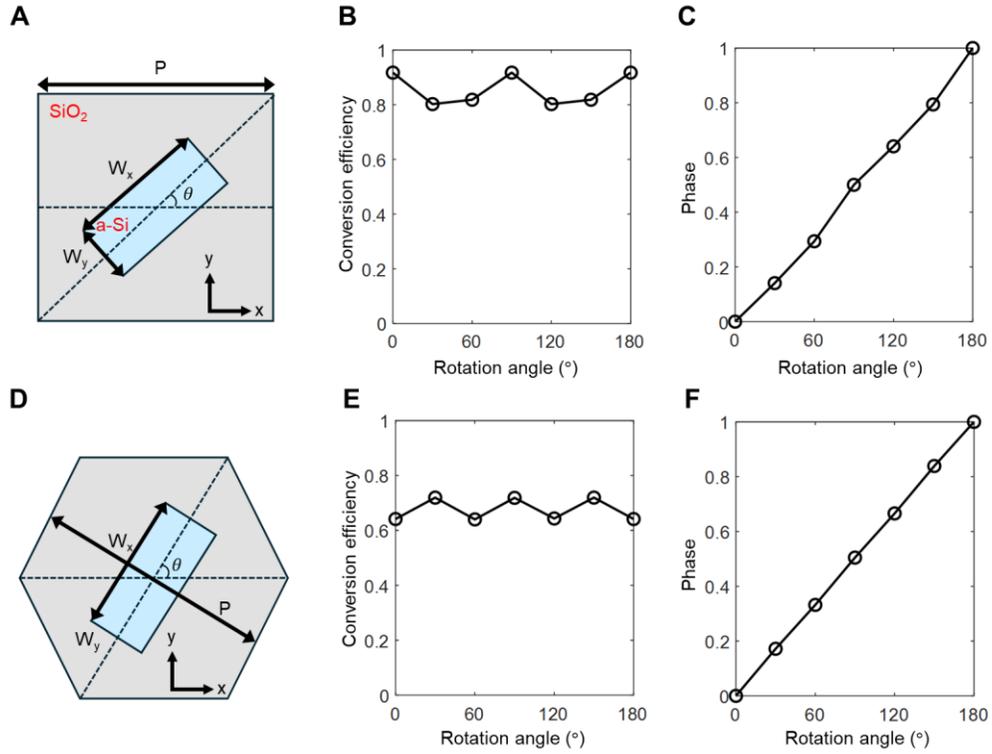

**Fig. S13. Meta-atoms for visible metalens.** The meta-atom for metalens in visible wavelength. (**A, D**) Schematic representation of the unit cell for square lattice and hexagonal lattice, respectively. P = 500 nm, $W_x$ = 400 nm, $W_y$ = 150 nm, and the height is 700 nm. The refractive index of amorphous silicon (a-Si) is 2.42 at the target wavelength of 632 nm. (**B, C, E, F**) The graphs depicting the conversion efficiency, and geometric phase as a function of the rotation angle ($\theta$) obtained by RCWA.

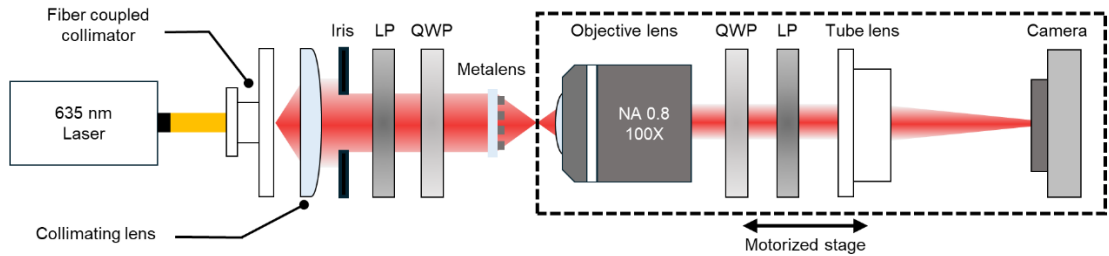

**Fig. S14. Z-scan imaging setup.** The schematic illustration of the optical setup for measuring the intensity distribution of propagating light after transmitting the metalens. The laser beam is collimated by a single-mode fiber-coupled collimator and collimating lens. It becomes left circularly polarized after passing through the LP and QWP. An objective (100× magnification, NA = 0.8) paired with a tube lens is used to image the intensity distribution of the laser beam after transmitting the metalens on the sCMOS camera. To filter out the $0^{th}$-order light, QWP and LP are placed after the objective lens. We mount the objective lens, QWP, LP, tube lens, and sCMOS camera on the motorized stage to observe the aliasing pattern radiating from the metalens.

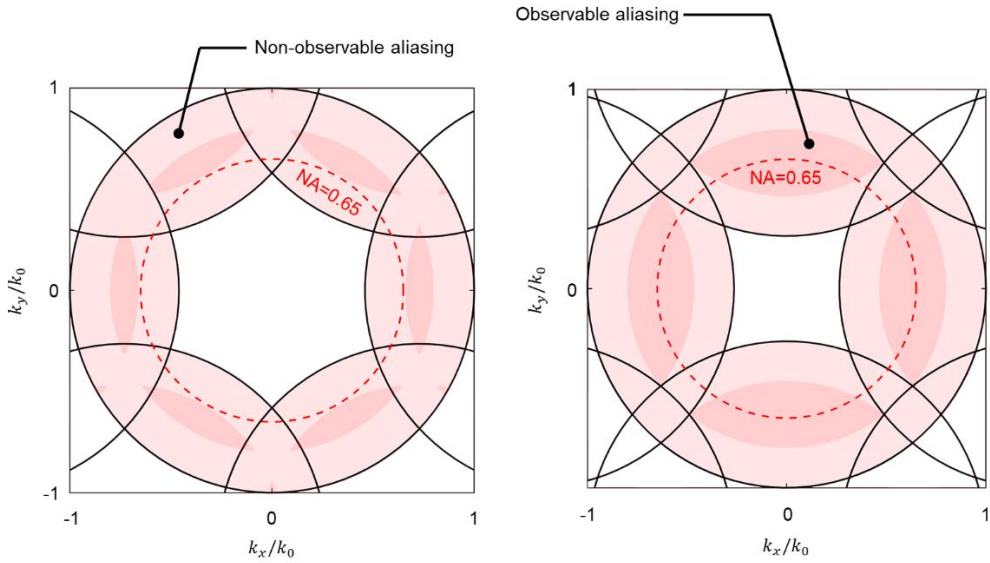

**Fig. S15. Observable aliasing spectral region restricted by an objective lens.** The representation of the observable aliasing spectral region, restricted by the objective lens (NA = 0.8) for a hexagonal sampling lattice (left) and a square sampling lattice (right). This is attributed to the fact that the propagating light having a NA larger than the objective lens NA cannot reach the camera. The red dotted line within the light cone denotes the NA of metalens. The aliasing occurring in the hexagonal sampling lattice cannot be observed since the aliasing spectrum for metalens (NA = 0.65) does not belong to the observable aliasing spectrum. Whereas the partial spectrum of aliasing in the square lattice can be observed.

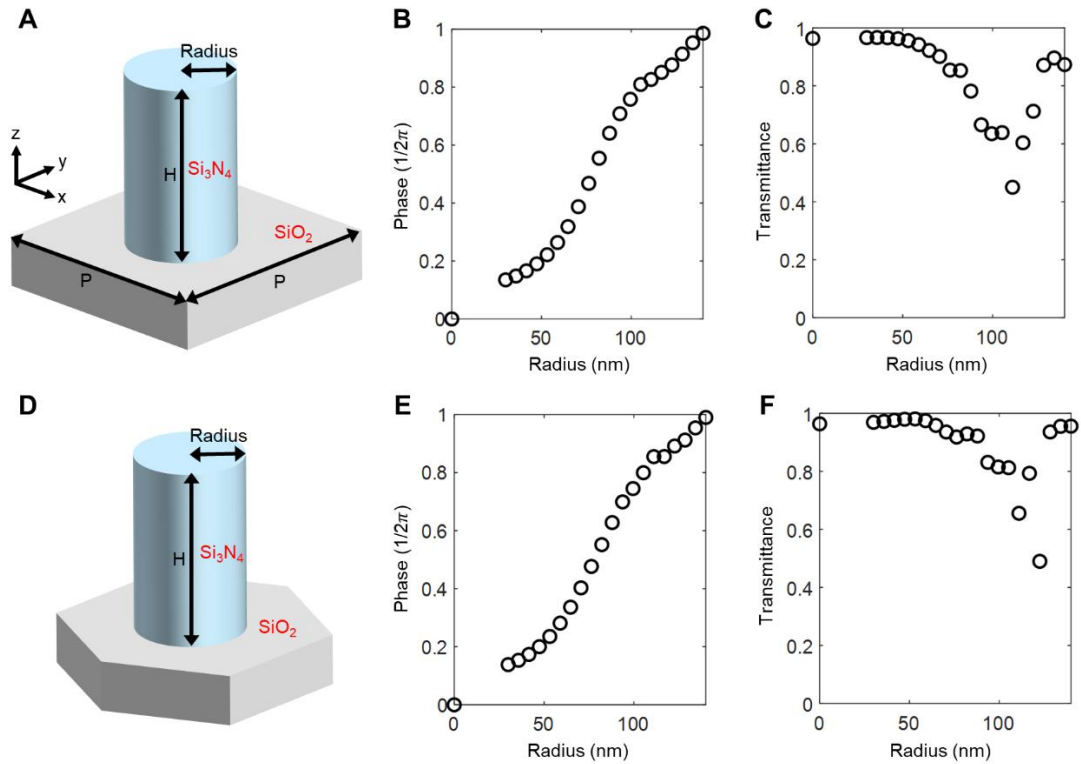

**Fig. S16. Phase and transmission results of meta-atoms in square lattice and hexagonal lattice.** (**A, D**) Cross-section view of meta-atoms in square lattice and hexagonal lattice. The height and periodicity (P) of the unit cell are 450 nm and 320 nm, respectively. (**B, C, E, F**) Graphs for the propagation phase and transmission data for meta-atoms at a wavelength of 405 nm, as calculated using RCWA. The range of radius for the meta-atom employed is between 30 nm and 140 nm, including an empty structure (*i.e.* radius = 0).

| θ°   | w$_x$ (nm) | w$_y$ (nm) | r (nm) | g (nm) | T$_1$ | T$_0$ | T$_{-1}$ | R |
|------|------|------|------|------|-------|-------|----------|-------|
| 19.7 | 164  | 180  | 80   | 80   | 0.540 | 0.030 | 0.208    | 0.142 |
| 23.9 | 164  | 148  | 80   | 67   | 0.637 | 0.028 | 0.086    | 0.066 |
| 27.4 | 148  | 180  | 80   | 67   | 0.697 | 0.018 | 0.025    | 0.100 |
| 30.4 | 164  | 180  | 80   | 67   | 0.705 | 0.001 | 0.007    | 0.278 |
| 34.2 | 148  | 180  | 53   | 67   | 0.751 | 0.005 | 0.053    | 0.180 |
| 36.6 | 148  | 180  | 40   | 67   | 0.741 | 0.004 | 0.085    | 0.160 |
| 39.3 | 148  | 180  | 40   | 67   | 0.797 | 0.015 | 0.105    | 0.077 |
| 42.5 | 164  | 180  | 40   | 67   | 0.811 | 0.028 | 0.090    | 0.070 |
| 46.3 | 148  | 180  | 40   | 67   | 0.796 | 0.062 | 0.080    | 0.047 |
| 51.2 | 148  | 180  | 40   | 67   | 0.764 | 0.118 | 0.075    | 0.031 |
| 54.1 | 148  | 180  | 40   | 67   | 0.725 | 0.152 | 0.075    | 0.036 |
| 57.5 | 148  | 180  | 40   | 67   | 0.683 | 0.182 | 0.074    | 0.042 |
| 61.7 | 148  | 180  | 40   | 67   | 0.616 | 0.229 | 0.088    | 0.055 |
| 67.0 | 148  | 180  | 40   | 67   | 0.536 | 0.270 | 0.102    | 0.078 |
| 74.6 | 148  | 180  | 40   | 53   | 0.476 | 0.178 | 0.188    | 0.107 |

**Table S1. Optimized parameters of meta-dimer.**

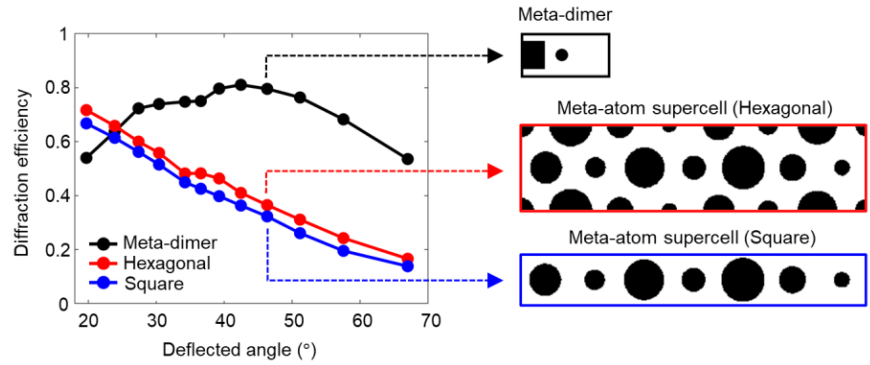

**Fig. S17. Beam steering efficiencies for metasurface and meta-dimer.** Illustration of the calculated diffraction efficiencies for metasurfaces and meta-dimers at various steering angles. Calculation is conducted using RCWA. The computational approach for calculating efficiency of the metasurfaces is detailed in Materials and Methods section.

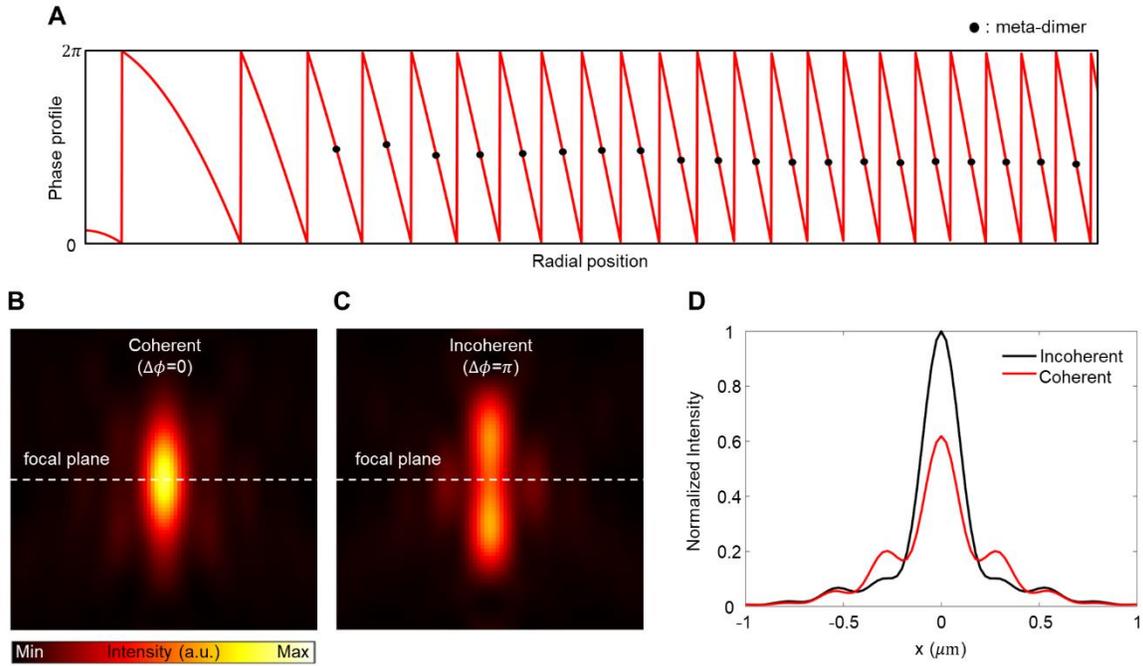

**Fig. S18. Coherent integration of meta-dimer with metasurface.** (**A**) The phase profile of the integrated lattice metalens (red solid line). The black dots are the phases of the desired diffracted order from each meta-dimer, which are shown to be coherently combined with the meta-atom part. (**B-D**) Simulated results for the two cases (B) where the meta-atom and meta-dimer array are coherently combined, and (C) when they are combined with a phase difference of π. (D) When combined incoherently, there is a reduction in peak intensity due to destructive interference at the focal point, leading to an increase of side lobes and a decrease in focusing efficiency (*46*).

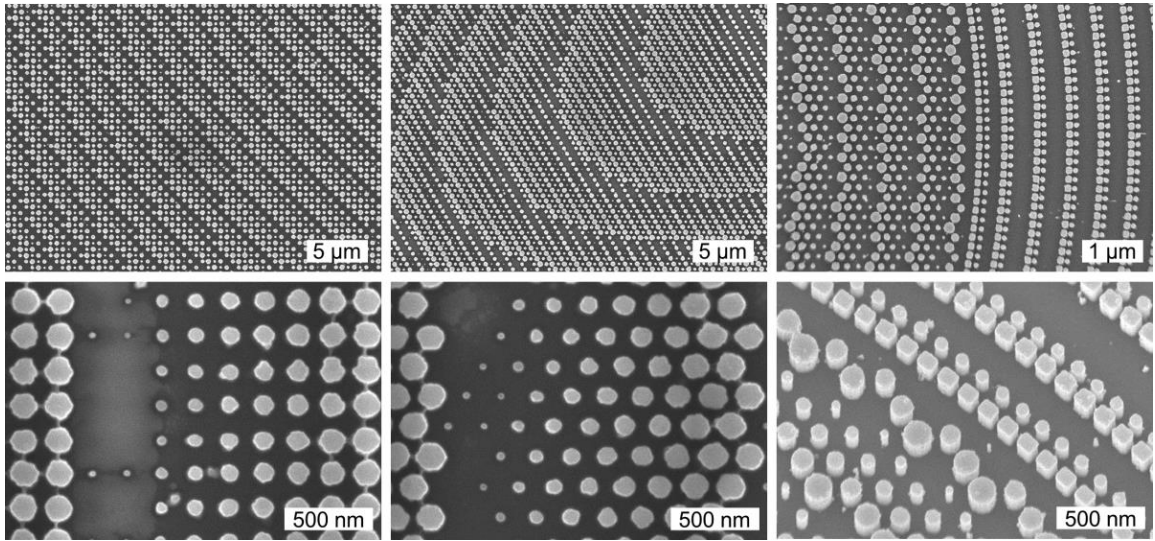

**Fig. S19. SEM images of UV metalenses.** The SEM images for square lattice metalens (left), hexagonal lattice metalens (middle), and integrated lattice metalens (right).

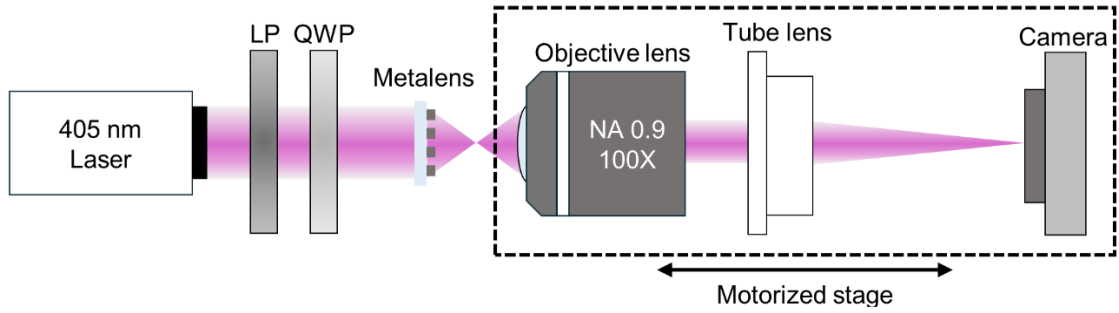

**Fig. S20. UV metalens measurement setup.** The schematic illustration of the optical setup for measuring the intensity distribution of propagating light after transmitting the metalens. The laser beam is left circularly polarized after passing through the LP and QWP. An objective (100× magnification, NA = 0.9) paired with a tube lens is used to image the intensity distribution of the laser beam after transmitting the metalens on the sCMOS camera. We mount the objective lens with a paired tube lens and the sCMOS camera on the motorized stage to investigate the aliasing of the metalens.

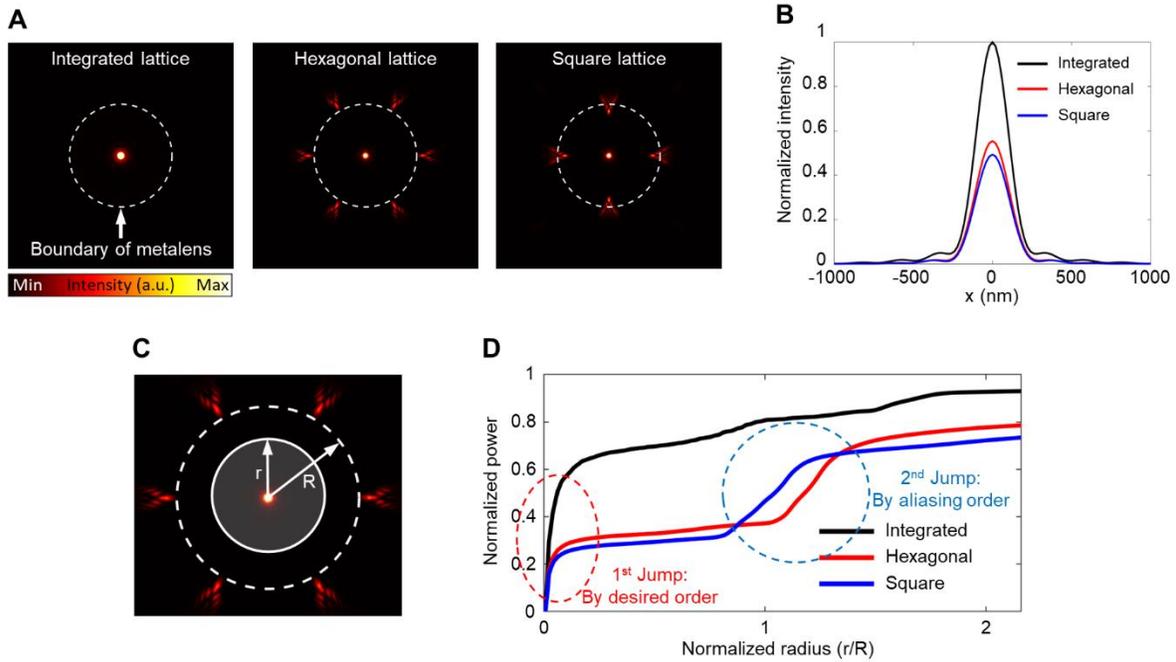

**Fig. S21. Simulation results for UV metalenses.** (**A**) Calculated intensity distributions at the focal plane for integrated lattice metalens and conventional metalenses. The white dotted circle represents the diameter of metalens at the sample plane. (**B**) The simulated point spread functions (PSF) for integrated lattice metalens and two conventional metalenses. (**C**) The schematic diagram for calculating normalized power. (**D**) Normalized incoming power within the circle encircling the focal point. The power is normalized by input power through each metalens. Therefore, as the radius increases, the normalized power tends to converge to the transmittance of the metalens. R is the radius of metalens and r is the radius of the encircling circle capturing the incoming power. In the vicinity of the focal point, a pronounced jump in incoming power is observed due to the focused beam, which is the desired diffraction. For the conventional metalens, a second jump occurs near the lens boundary (r/R = 1) since the power of the aliasing order starts to be counted. However, in the case of the integrated lattice metalens, a significant reduction of the second jump can be observed. Additionally, in square lattice metalenses, a jump due to the aliasing order occurs earlier compared to hexagonal lattice metalenses. This is attributed to the relatively wider aliasing spectrum in square lattices, resulting in relatively smaller diffraction angles for aliasing diffraction orders compared to those in hexagonal lattices.

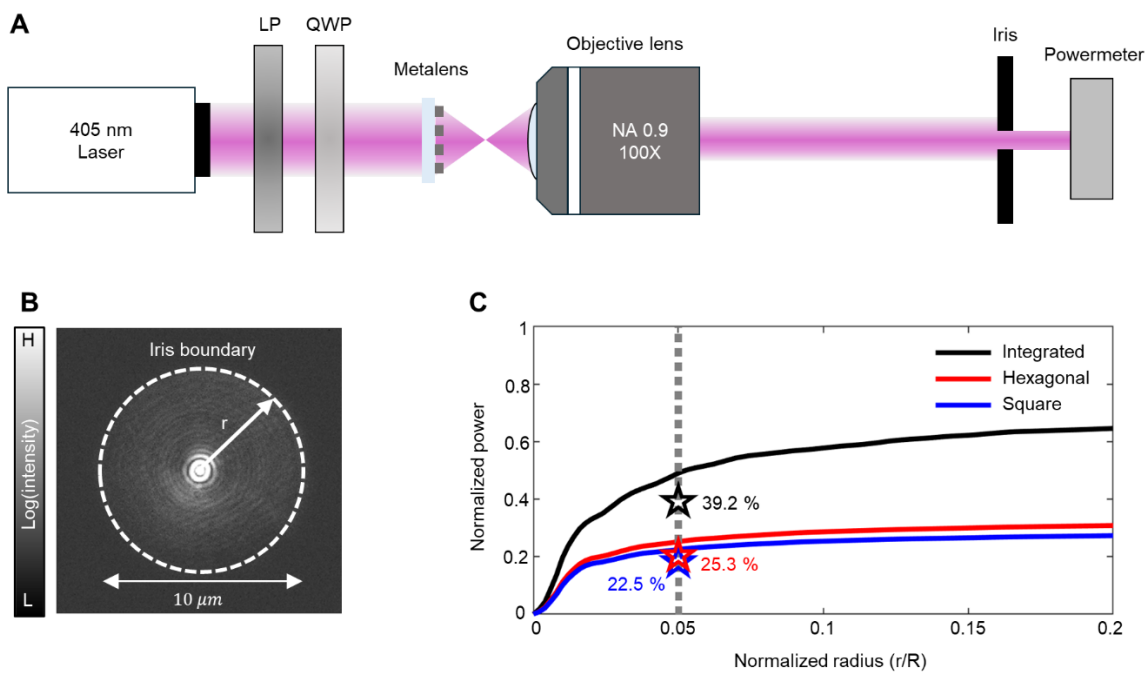

**Fig. S22. Measured focusing efficiency of UV metalenses.** (**A**) Schematic illustration of the optical setup for measuring the focusing efficiency of UV metalenses. The focusing efficiency of metalenses is measured using a powermeter. (**B**) The image of the intensity distribution of the focused light entering the powermeter, which is captured by the sCMOS camera. The area beyond 5 μm from the focusing point is blocked by the iris, whose normalized radius (r/R) corresponds to 0.05. (**C**) Results of the measured focusing efficiency for three metalenses and simulated normalized power. Pentagram signs represent the measured focusing efficiency of metalenses. The measured (simulated) efficiency for integrated lattice, hexagonal lattice, and square lattice metalenses at r/R = 0.05 are 39.2% (49.1%), 20.3% (25.3%), and 18.2 % (22.5%), respectively.

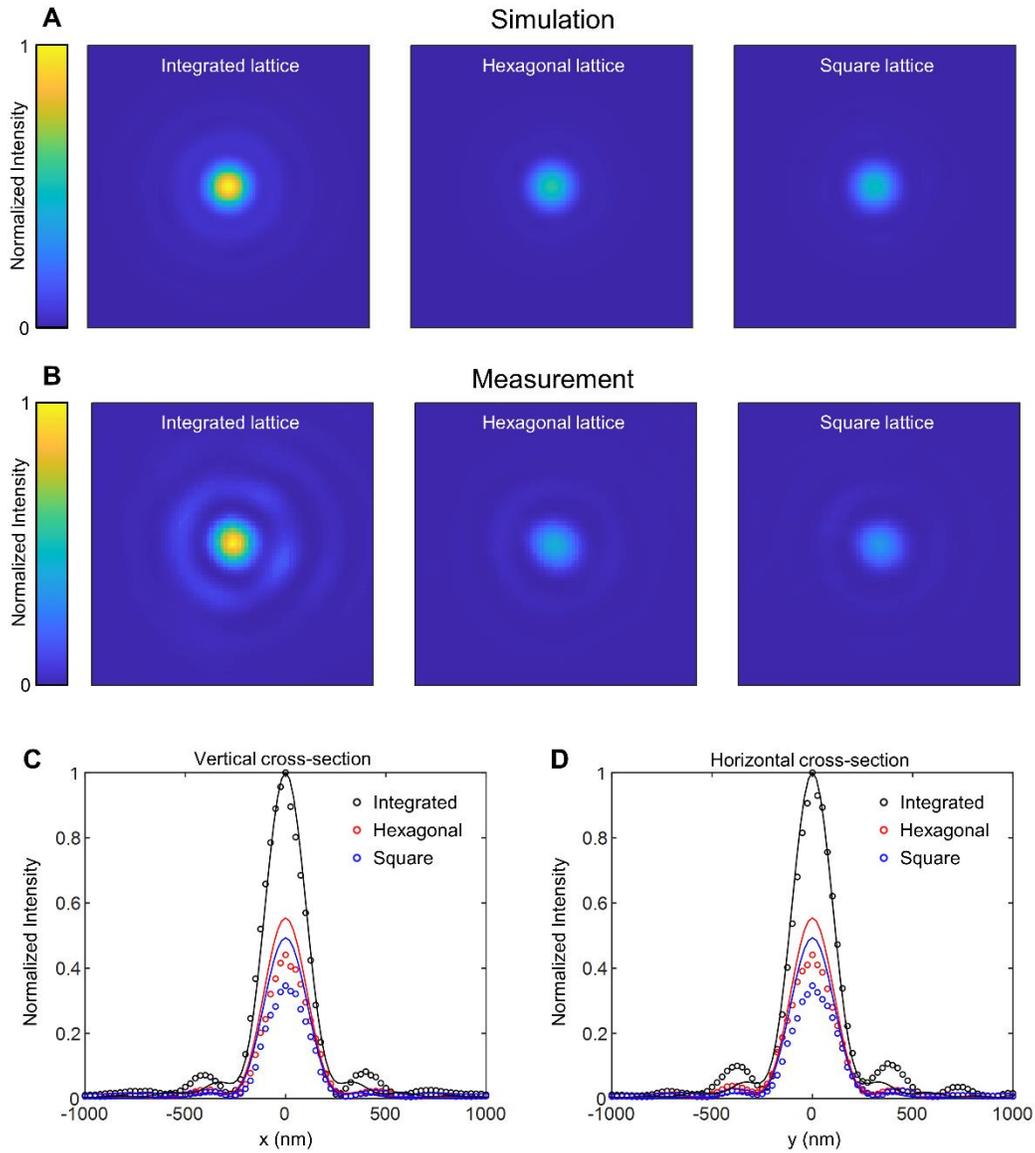

**Fig. S23. Simulated and measured PSF of UV metalenses.** (**A**) Simulated, and (**B**) measured PSF for integrated lattice (left), hexagonal lattice (middle), and square lattice (right) metalens, respectively. Each PSF is normalized by the maximum intensity value in PSF for integrated lattice metalens. (**C, D**) Vertical and horizontal cross-sectional view of PSF data. Dots (solid lines) represent measured (simulated) PSF data. Each measured (simulated) full width at half maximum (FWHM) of PSFs is 227 nm (233 nm), 260 nm (244 nm), and 262 nm (245 nm), respectively.